\documentclass[]{elsarticle}

\usepackage{natbib} 
\usepackage{subfigure}
\usepackage{graphicx}
\usepackage{color}
\usepackage{lastpage}

\usepackage[T1]{fontenc}
\usepackage[utf8]{inputenc}
\usepackage{amsmath}
\usepackage{amssymb}
\usepackage{bm}

\usepackage{soul}

\newcommand{\R}{\sqrt{f-t_0}}
\newcommand{\xio}{x^\mathrm{in,out}}
\newcommand{\thio}{\theta^\mathrm{in,out}}
\newcommand{\Mio}{m^\mathrm{in,out}}
\newcommand{\e}{\epsilon}
\newcommand{\xin}{x^\mathrm{in}}
\newcommand{\xout}{x^\mathrm{out}}
\newcommand{\thin}{\theta^\mathrm{in}}
\newcommand{\thout}{\theta^\mathrm{out}}
\newcommand{\Min}{m^\mathrm{in}}
\newcommand{\Mout}{m^\mathrm{out}}
\newcommand{\hs}{\hat{s}}
\renewcommand{\d}{\mathrm{d}}
\newcommand{\glv}{\gamma_\mathrm{LV}}
\newcommand{\gsl}{\gamma_\mathrm{SL}}
\newcommand{\gsv}{\gamma_\mathrm{SV}}
\newcommand{\sa}{L/2-\Sigma}
\renewcommand{\sb}{L/2+\Sigma}

\graphicspath{{./figs/}}

\begin{document}

\begin{frontmatter}

\title{Coiling of an elastic beam inside a disk: a model for spider-capture silk}

\author[idaCNRS,idaUPMC]{Herv\'e Elettro} 
\author[oxf]{Fritz Vollrath} 

\author[idaCNRS,idaUPMC]{Arnaud Antkowiak} 
\author[idaCNRS,idaUPMC]{S\'ebastien Neukirch} \ead{sebastien.neukirch@upmc.fr}

\address[idaCNRS]{Centre National de la Recherche Scientifique, UMR 7190, Institut Jean Le Rond d’Alembert, F-75005 Paris, France}
\address[idaUPMC]{Sorbonne Universit\'es, UPMC Univ Paris 06, UMR 7190, Institut Jean Le Rond d’Alembert, F-75005 Paris, France}
\address[oxf]{Oxford Silk Group, Zoology Department, University of Oxford, UK}

\date{\today}

\begin{abstract}
Motivated by recent experimental observations of capillary-induced spooling of fibers inside droplets both in spider capture silk and in synthetic systems, we investigate the behavior of a fiber packed in a drop. Using a simplified 2D model, we provide analytical predictions for the buckling threshold and the deep post-buckling asymptotic behavior. The threshold for spooling is found to be in particularly good agreement with experimental results. We further solve the Elastica equations for a fiber confined in a soft potential, and track the equilibrium paths using numerical continuation techniques. A wealth of different paths corresponding to different symmetries is uncovered, and their stability is finally discussed.
\end{abstract}

\end{frontmatter}


%
%
%
%
\section{Introduction} \label{sec:level1}
\begin{figure}[ht]
\begin{center}
\includegraphics[height=2.5cm]{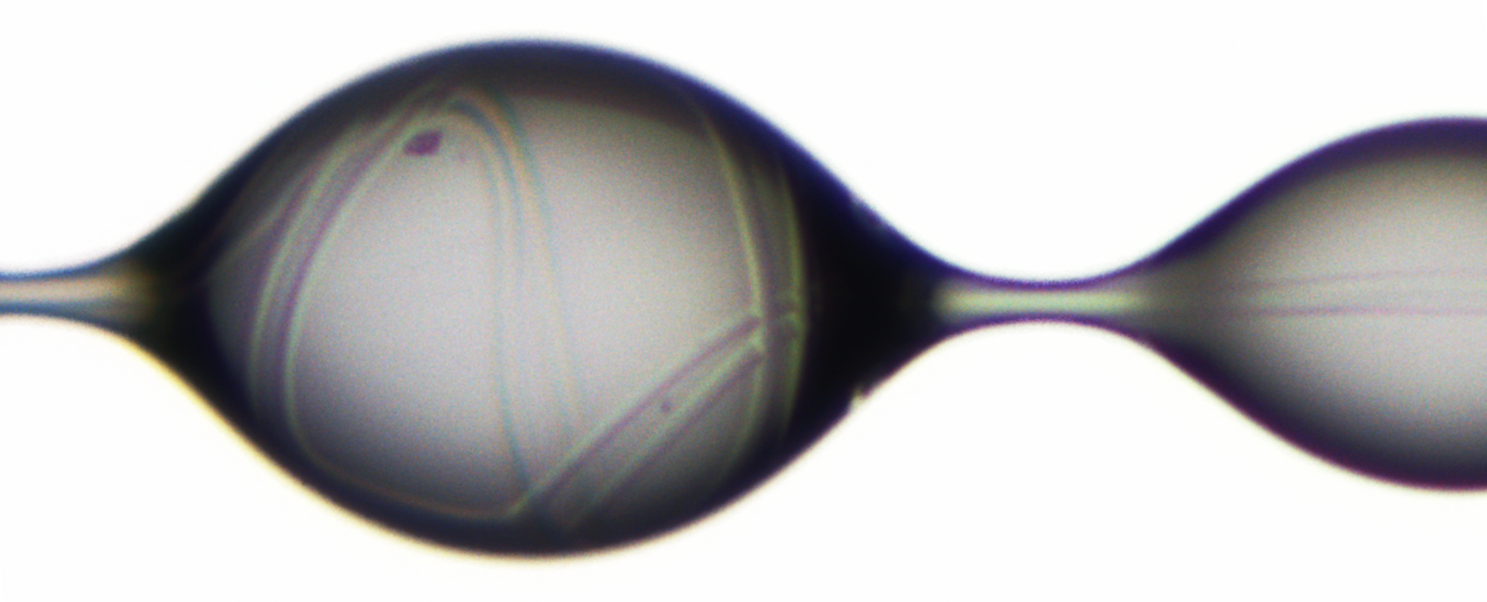}\hspace{1cm}\includegraphics[height=2.5cm]{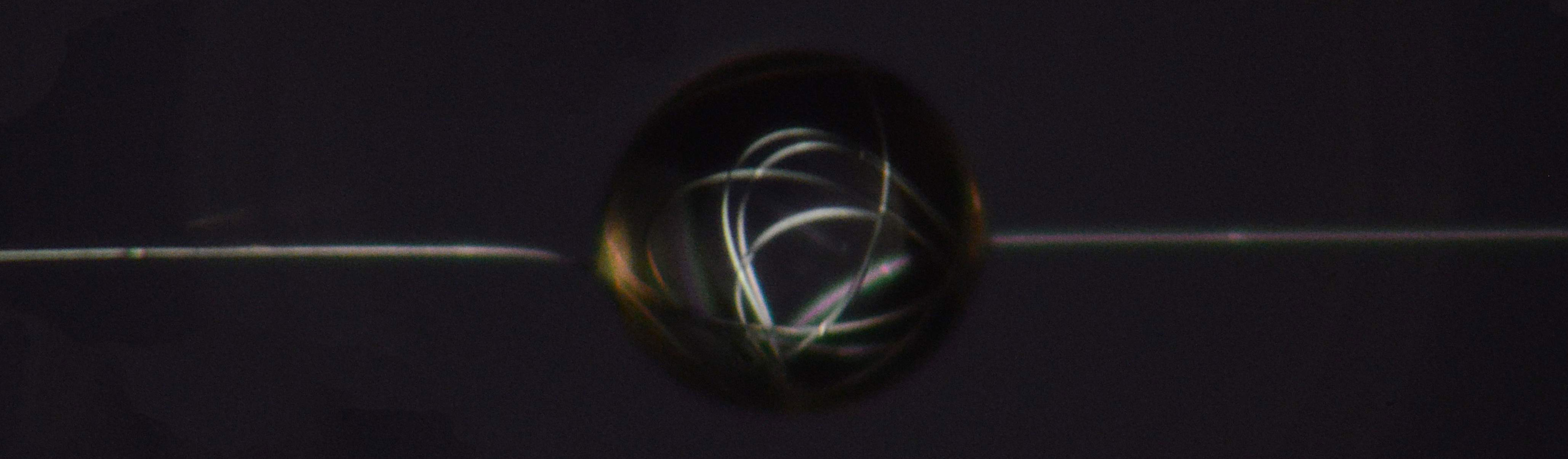}
\caption{Experiments on fibers bent inside liquid drops. Left: microscopic photograph of spider capture silk. Flagelliform core filaments are seen to be coiled and packed inside a (typically 300 $\mu$m wide) glue droplet. Right: same mechanism reproduced artificially with a 200 $\mu$m synthetic droplet and fiber (see experimental verification section in section \ref{section:exp}). Reproduced from \cite{Elettro2015In-drop-capillary}.}
\label{fig:intro}
\end{center}
\end{figure}

%
%
%
The mechanical properties of spider silk are often presented as outstanding \cite{Foelix2010,Omenetto2010}. An indeed, most silk threads outperform the best man-made fibers, such a Kevlar, at least in terms of toughness \cite{Vollrath2001}. To a large extent, these properties rely on the molecular architecture of the silk. For example, it has been shown that the building blocks of flagelliform silk involve molecular nanosprings \cite{Becker2003a}. In 1989 however, a team comprising a zoologist and a physicist  reported on coiling and packing of the core filament inside a glue droplet \cite{Vollrath1989}. This {\em windlass mechanism}, as it was called, provided indirect evidence that the glue droplets may as well play a role in the mechanical response of the silk thread. These results have been a subject of debate in the community, and it is only very recently that the mechanism has been observed to be active in a real spider web, see Fig.~\ref{fig:intro}-Left and \cite{Elettro2015In-drop-capillary}. A natural question that arises in this context is the role played by the molecular structure of the silk and the glue in the observed coiling. An experimental answer to this question is provided in Fig.~\ref{fig:intro}-Right, where a micron-sized artificial thread bearing a silicon oil droplet also exhibits the coiling mechanism and packing behavior, therefore demonstrating that capillarity and elasticity are sufficient ingredients to explain the mechanism.

Interestingly, the shape adopted by the filament inside the drop can be as different as a perfectly ordered closely-packed annular bundle or a completely disordered tangle. This behavior is reminiscent of the organization of packed wires in rigid  \cite{Stoop2011} and elastic \cite{Vetter2014} spherical shells, patterns of folded structures such as plant leaves or crumpled paper \cite{Boue:2006,Couturier2011}, and DNA packing inside capsids \cite{lamarque:2004,Klug2005, Leforestier2009}. The purpose of the present paper is to explore theoretically in a simplified setting the shape and stability of strongly post-buckled states in order to lay down the basis for a deeper understanding of the windlass mechanism.

The paper is organized as follows. 
In section \ref{sec:model} we present the problem and the equilibrium equations. 
In section \ref{section:buckling} we perform a linear stability analysis of the straight beam and  predict the buckling threshold.
Experimental results are confronted to theoretical in section \ref{section:exp} experiments.
Finally, we describe the non-linear response of the system in terms of equilibrium solutions and their stability in section \ref{sec:post-buckling}
\section{Model} \label{sec:model}
%
%
%
%
\begin{figure}[ht]
\begin{center}
\includegraphics[width=.6\columnwidth]{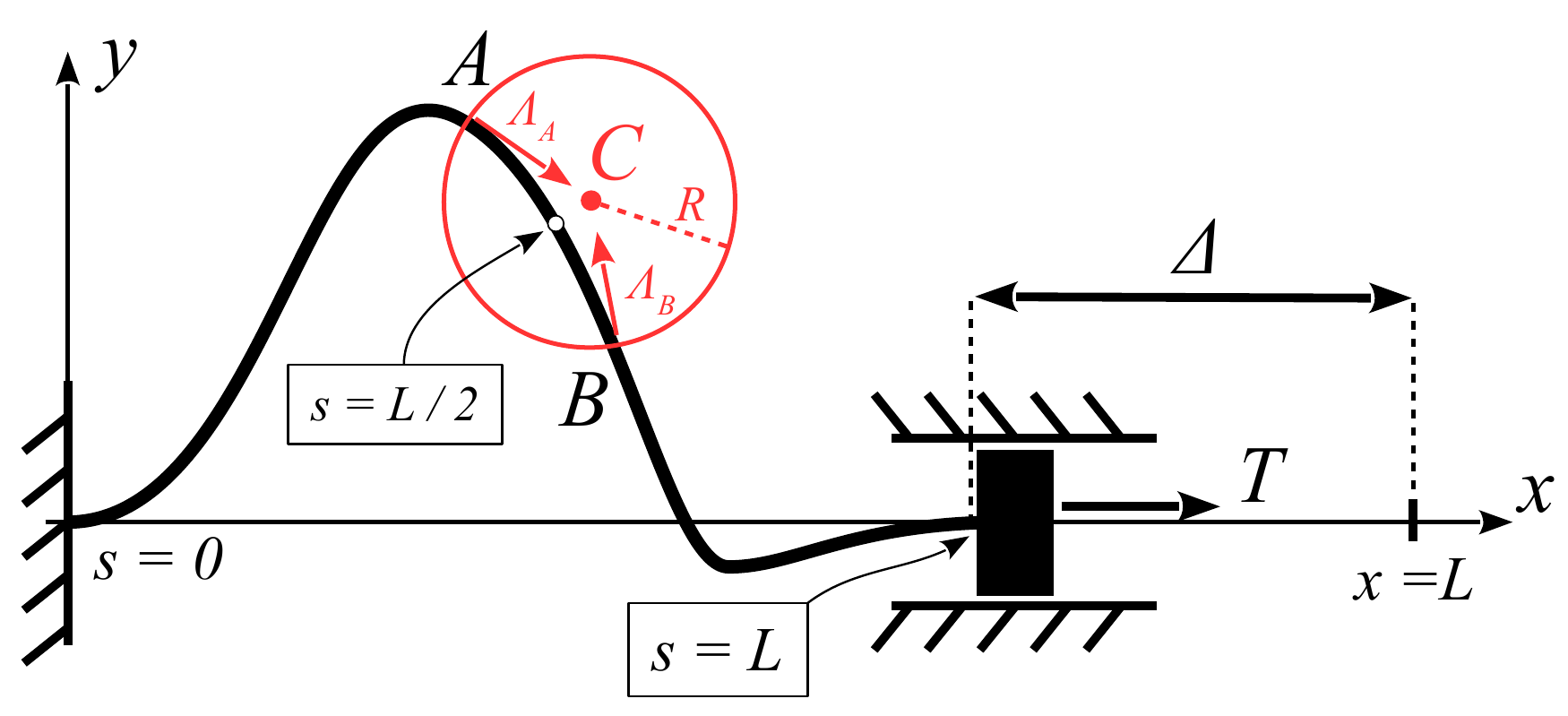}
\caption{An elastic beam held in tension at its extremities, and buckling under the action of compressive forces at a disk. The beam is clamped at both ends. The deformation of the beam is described by the angle $\theta(S)$ between the tangent to the beam and the $x$-axis, where $S \in \left[ 0, L \right]$ is the arc-length along the beam.}
\label{fig:notations}
\end{center}
\end{figure}
We consider an elastic beam in interaction with a liquid disk and under the action of a tensile end-load.
As indicated in Fig.~\ref{fig:notations}, we restrict to planar deformations of the beam, $X$ and $Y$ denoting the horizontal and vertical directions respectively. 
The beam has length $L$ and a circular cross-section of radius $h$. We work under the slender ($L\gg h$) Euler-Bernoulli hypotheses where the beam is considered inextensible and unshearable. Configurations are thus fully described by the position and orientation of the centerline.
We use the arc-length $S \in [0,L]$ and note $\theta(S)$ the angle between the tangent of the beam and the horizontal.
The presence of the liquid disk generates capillary forces due to the contrast of surface energy, the interaction energy of the beam with the liquid being smaller than the interaction energy of the beam with the air. Capillary forces are two-fold: $(i)$ meniscus forces applied on the beam at the entrance and exit of the disk, and $(ii)$ barrier forces that prevent the beam from exiting the disk elsewhere than at the meniscus points. We consider that the drop is undeformable and thus remains a disk throughout the experiments.
As shown in \ref{section:varia}, meniscus forces are pointing toward the center of the disk (see Equations (\ref{sys:app-saut-forces-x}) and (\ref{sys:app-saut-forces-y})) and their intensity is related to the angle between their direction and the tangent to the beam at the meniscus points (see Equation (\ref{equa:app-intensity-jumps})).
A soft-wall barrier potential \cite{ManningBulman:2005}
\begin{equation}
V(X,Y) = \frac{V_0}{1 + \rho - (1/R) \sqrt{(X-X_C)^2 + (Y-Y_C)^2}   }
\end{equation}
is used to retain the beam inside the disk, centered on $(X_C,Y_C)$ and of radius $R$. The small dimensionless parameter $\rho$ is introduced to avoid the potential to diverge at the meniscus points $A$ and $B$, where the rod enters and exits the disk.
The intensity $V_0$ of the potential is chosen to be small, the hard-wall limit being $V_0 \to 0$.
Kinematics, relating the position $(X,Y)$ of the rod and the inclination $\theta$ of its tangent $(\cos \theta,\sin \theta)$ with the horizontal, the bending constitutive relation, relating the curvature $\theta'(S)$ to the moment $M(S)$, and finally force $(N_x,N_y)$ and moment balance are detailed in \ref{section:varia} and read
\begin{subequations}
\label{sys:equilibre}
\begin{align}
X'(S)= \cos \theta \quad &, \quad  Y'(S)= \sin \theta \\
 EI \, \theta'(S) = M \quad &, \quad  M'(S) = N_x \sin \theta - N_y \cos \theta \\ 
N_x'(S) &= \chi  \, \frac{\partial V}{\partial X} + \delta(S-S_A) \, \Lambda_A \, \frac{ X_A - X_C }{R}
+ \delta(S-S_B) \, \Lambda_B \, \frac{ X_B - X_C }{R}\\ 
N_y'(S) &= \chi \, \frac{\partial V}{\partial Y} + \delta(S-S_A) \, \Lambda_A \,  \frac{ Y_A - Y_C }{R}
+ \delta(S-S_B) \, \Lambda_B \, \frac{ Y_B - Y_C }{R}
\end{align}
\end{subequations}
%
%
where $S$ is the arc-length along the rod, and $()' = \mathrm{d}()/\mathrm{d} S$.
We define the coordinates of point $A$ as $(X_A,Y_A)=(X(S_A),Y(S_A))$, same for point $B$. 
Note that the potential $V$ has the dimension of an energy per unit of arc-length of the beam.
For $S \in \left[S_A ; S_B \right]$ the rod lies inside the disk and we have $\chi=1$, otherwise $\chi=0$. The Dirac distribution $\delta(S)$ localizes meniscus forces at points $A$ and $B$. The rod material has Young's modulus $E$ and the second moment of area $I=\pi \, h^4 /4$.
The intensities $\Lambda_A$ and $\Lambda_B$ of the meniscus forces are unknown but related to surface tension $\glv$ through Equation (\ref{equa:app-intensity-jumps}), where $F_\gamma = 2 \pi h \, \glv \cos \alpha_Y$ with $\alpha_Y$ being the Young-Dupr\'e wetting angle ($\gsv - \gsl=\glv \cos \alpha_Y$), and where $V_A=V_B=V_0/\rho$ are small compared to $F_\gamma$.
We restrict ourself to cases where the disk is centered on the mid-point of the rod, that is we introduce $\Sigma$ such that $S_A = \sa$ and $S_B=\sb$. The rod has then $2 \, \Sigma$ of its arc-length spent inside the disk. Finally the external applied tension is noted $T=N_x(L)$.
\subsection*{Non-dimensionalization} 

We use the diameter $D=2R$ of the disk as unit length, and the buckling load $EI/D^2$ as unit force. We thus introduce the following dimensionless quantities
\begin{subequations}
\label{eq:adim}
\begin{align}
s = \frac{S}{D} \; ; \; \sigma = \frac{\Sigma}{D} \; ; \; \ell = \frac{L}{D} \; ; \; 
(x,y) = \frac{(X,Y)}{D} \; ; \;  n = \frac{N D^2}{EI} \; ; \; t = \frac{T \, D^2}{EI}  \\
f_\gamma = \frac{F_\gamma \, D^2}{EI}  \; ; \; m = \frac{MD}{EI}   \; ; \; 
\lambda_{A,B} = \frac{\Lambda_{A,B} \, D^2}{EI} \; ; \; (v,v_0) = \frac{(V,V_0) \,D^2}{EI} 
\end{align}
\end{subequations}
and $\delta(s)=D\, \delta(S)$.
We then have $v(x,y) = v_0 \, \left(1 + \rho - 2 \, \sqrt{(x-x_C)^2 + (y-y_C)^2} \right)^{-1}$ and
\begin{subequations}
\label{sys:equilibre-adim}
\begin{align}
x'(s)= \cos \theta \quad &, \quad  y'(s)= \sin \theta \\
\theta'(s) = m \quad &, \quad m'(s) = n_x \sin \theta - n_y \cos \theta \\ 
n_x'(s) &= \chi  \, \frac{\partial v}{\partial x} + 2 \delta(s-s_A) \, \lambda_A \, ( x_A - x_C )
+ 2 \delta(s-s_B) \, \lambda_B \, ( x_B - x_C) \label{equa:bal-foce-x} \\ 
n_y'(s) &= \chi \, \frac{\partial v}{\partial y} + 2 \delta(s-s_A) \, \lambda_A \,  ( y_A - y_C )
+ 2 \delta(s-s_B) \, \lambda_B \, (y_B - y_C) \label{equa:bal-foce-y}
\end{align}
\end{subequations}
where $()' = \d() / \d s$, and $s_A = \ell/2 - \sigma$, $s_B = \ell/2 + \sigma$.

\subsection*{Boundary-value problem} \label{section:bvp}
We consider $v_0$, $\rho$, $f_\gamma$, and $\ell$ as fixed parameters and we look for equilibrium solutions by integrating equations (\ref{sys:equilibre-adim}) with the initial conditions
\begin{equation}
x(0)=0 \; ; \; y(0)=0 \; ; \; \theta(0)=0 \; ; \; m(0)=m_0 \; ; \; n_x(0)=n_{x0} \; ; \; n_y(0)=n_{y0}
\label{equa:initcnd}
\end{equation}
where $m_0$, $n_{x0}$, and $n_{y0}$ are unknowns to be accompanied with $\sigma$, $x_C$, $y_C$, $\lambda_A$, and $\lambda_B$. We therefore have 8 unknowns which are balanced by the following 7 conditions.
At the $s=\ell$ end of the rod, clamped boundary conditions read
\begin{equation}
y(\ell)=0 \quad ; \quad \theta(\ell)=0
\label{equa:finalbcnd}
\end{equation}
The requirement that points $A$ and $B$ lie on the circle yields the conditions
\begin{equation}
\left[ x_A - x_C \right]^2 + [ y_A - y_C]^2 = 1/4 \quad ; \quad
[ x_B - x_C ]^2 + [ y_B - y_C]^2 = 1/4
\label{equa:AB-cercle}
\end{equation}
and the three force balances related to forces coming from the disk read
\begin{subequations}
\label{equa:sauts-force}
\begin{align}
n_x(s_A^-) &= n_x(s_B^+) \\
n_y(s_A^-) &= n_y(s_B^+)  \\
- 2 f_\gamma + v_A + v_B 
&- 2 \lambda_A  \left[
( x_A - x_C )  \, \cos \theta(s_A)+   
( y_A - y_C )  \, \sin \theta(s_A)
 \right] \nonumber \\
&+ 2 \lambda_B \left[
( x_B - x_C )  \, \cos \theta(s_B)+   
( y_B - y_C )  \, \sin \theta(s_B)   
 \right]  =0 \label{8C}
\end{align}
\end{subequations}
The solution set is thus a $8-7=1$ dimensional manifold and we plot in Section \ref{sec:post-buckling} different solution paths for several values of the parameter $f_\gamma$.

\section{Buckling threshold} \label{section:buckling}
\begin{figure}[h]
\centering
\includegraphics[width=.48\columnwidth]{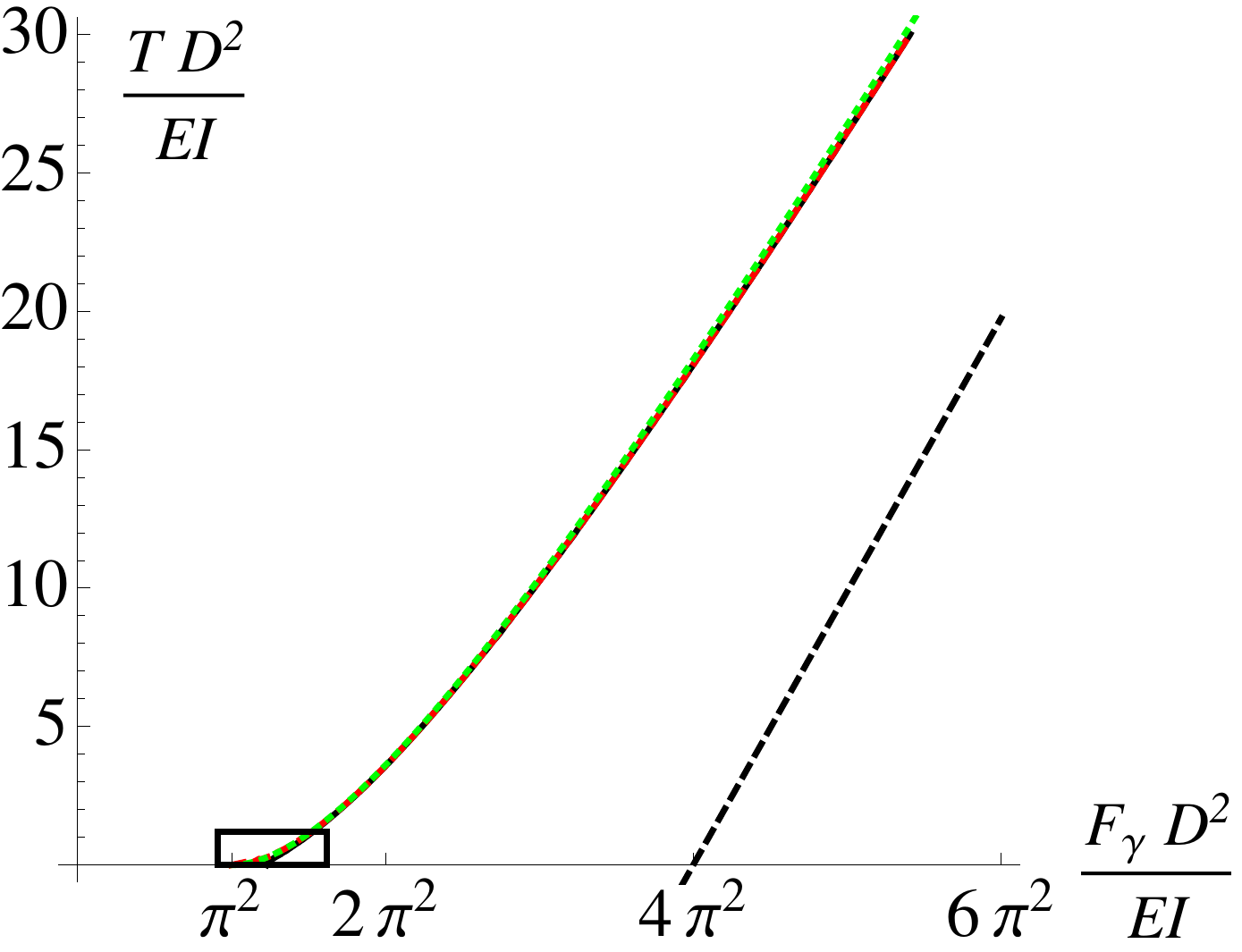}
\includegraphics[width=.41\columnwidth]{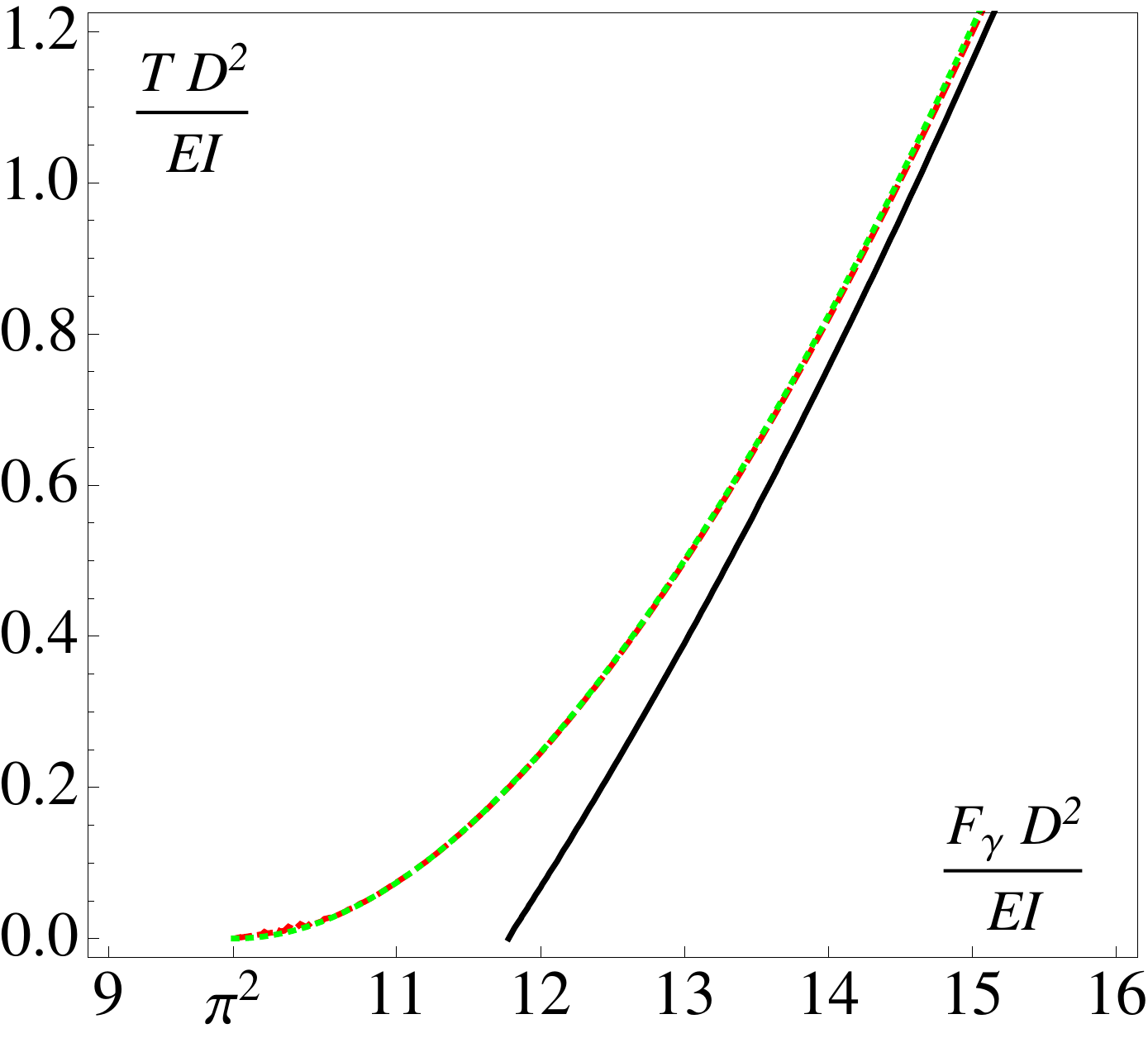}
\caption{Buckling curve for $\ell=5$. (Left) Curves defined by Equations (\ref{equa:buckling-curve}) (\ref{equa:buckling-curve-approx-1}) (\ref{equa:tbuck}). At this scale the three curves are almost indistinguishable. The asymptote $t=f_\gamma - 4 \pi^2$ is shown dashed.
(Right) Zoom corresponding to the rectangle shown on the Left. The approximations (\ref{equa:buckling-curve-approx-1}) (\ref{equa:tbuck}) are still hard to distinguish, but are seen to deviate from the exact curve (\ref{equa:buckling-curve}), shown continuous and black.}
\label{fig:bucklingcurve}
\end{figure}

%
%
%
%
%
%
The trivial solution $x(s)=s$, $y(s)=\theta(s)=m(s)=n_y(s)=0$ to Equations (\ref{sys:equilibre-adim}) with boundary conditions (\ref{equa:initcnd})-(\ref{equa:sauts-force}), exists for any value of the load $t$.
Nevertheless, for given values of the parameters $v_0$, $\rho$, $f_\gamma$, and $\ell$, there is a threshold value of $t$ under which the trivial solution ceases to be stable and buckling occurs.
We look for the first buckling mode which is symmetrical with respect to the axis joining the center of the disk $(x_C,y_C)$ and the beam midpoint $(x(\ell/2),y(\ell/2))$.
We linearize equations (\ref{sys:equilibre-adim}) for small deflections, $|y(s)| \sim \epsilon$, and small slopes, $|\theta(s)| \sim \epsilon$, with $ 0 < \epsilon \ll 1$, see \ref{section:weakly-non-lin} for a comprehensive exposition of this perturbation expansion.
As in the buckling configuration the rod has virtually no packing interaction with the disk, we set $v_0=0$.
The first four equations of system (\ref{sys:equilibre-adim}) become $x'(s)=1$, $y'(s)=\theta(s)$, $\theta'(s)=m(s)$, and $m'(s)=n_x(s) \,  \theta(s) - n_y(s)$. We then have $x_C=\ell/2$, $x_A=\ell/2  - \sigma$, and $x_B=\ell/2 + \sigma$.
At order $\epsilon^0$, equation (\ref{equa:AB-cercle}) yields $\sigma=1/2$. As symmetry imposes $y_A=y_B$ and $\theta(s_A)=-\theta(s_B)$, Equation (\ref{equa:AB-cercle}) at order $\epsilon^2$ imposes $y_C=y_A=y_B$.
Then (\ref{8C}) at order $\epsilon^0$ yields $2 f_\gamma = \lambda_A+\lambda_B$ and, as symmetry requires $\lambda_A =\lambda_B$, we finally obtain  $f_\gamma = \lambda_A=\lambda_B$. 
Following symmetry we introduce $\hs = s-\ell/2$.
The three functions $y(\hs)$, $\theta(\hs)$, and $m(\hs)$ are then respectively even, odd, and even functions of the variable $\hs$.
We focus on the right half of the system, $\hs \in [0; \ell/2]$.
Equation (\ref{equa:bal-foce-x}) is integrated to yield $n_x(\hs)=t-f_\gamma$ for $\hs \in [0 ; 1/2]$ and  $n_x(\hs)=t$ for $\hs \in [1/2 ; \ell/2]$.
Equation (\ref{equa:bal-foce-y}) shows that $n_y(\hs) = \mathrm{const.}$ and from $m'(\hs)=n_x(\hs) \,  \theta(\hs) - n_y(\hs)$ we see that $n_y(\hs)$ has to be odd, hence zero.
We finally arrive at the reduced system
\begin{align}
\theta'' &= - (f_\gamma-t) \, \theta  \mbox{~ for ~}   \hs \in [0 ; 1/2] & \\
\theta'' &= t \, \theta  \mbox{~ for ~}  \hs  \in [1/2 ; \ell/2] 
\end{align}
and we restrict to the $f_\gamma>t \geq 0$ case. Integrating these equations and using the boundary conditions $\theta(\hs = 0)=0=\theta(\hs = \ell/2)$ and the matching conditions $\theta(\hs= 1/2^-)=\theta(\hs=1/2^+)$ and $m(\hs=1/2^-)=m(\hs=1/2^+)$, we obtain the buckling condition
\begin{equation}
\sqrt{f_\gamma - t} \, \tanh \frac{(\ell-1) \, \sqrt{t}}{2}  + \sqrt{t} \, \tan \frac{\sqrt{f_\gamma-t}}{2} = 0
\label{equa:buckling-curve}
\end{equation}
which is plotted in Figure~\ref{fig:bucklingcurve}. The two interesting asymptotic limits of the curve defined by (\ref{equa:buckling-curve}) are $(i)$ if $t \to 0$ then $f_\gamma \to \pi^2 + 8/\ell$, and $(ii)$ if $f_\gamma \to +\infty$ then $t \to f_\gamma - 4 \pi^2$.

\subsection*{Approximations to the buckling load}
%
%
In the case where $\ell \gg 1$, we simplify Equation (\ref{equa:buckling-curve}) and find
\begin{equation}
\sqrt{f_\gamma - t}   + \sqrt{t} \, \tan \frac{\sqrt{f_\gamma-t}}{2} = 0
\label{equa:buckling-curve-approx-1}
\end{equation}
This formula has the same large $f_\gamma$ limit as (\ref{equa:buckling-curve}) and in fact as $\ell \to +\infty$, the curve defined by (\ref{equa:buckling-curve}) tends to the curve defined by (\ref{equa:buckling-curve-approx-1}) everywhere but in a boundary layer around $(f_\gamma,t)=(\pi^2,0)$. Indeed even if $\ell$ is large, for small $t$ the $\tanh$ term cannot be approximated by 1 if $t \sim 1/\ell^2$.
It is convenient to have an explicit formula $t=t(f_\gamma)$ for buckling and we introduce the approximation
\begin{equation}
t_b(f_\gamma) = f_\gamma - 4 \pi^2 - \frac{108 \pi^4}
{ 3 \pi^4 -40 \pi^2 + (3 \pi^2-28) \, f_\gamma + (32 \pi - 6 \pi^3) \, \sqrt{f_\gamma}}
\label{equa:tbuck}
\end{equation}
This last formula has the same behavior as (\ref{equa:buckling-curve-approx-1}) at low $t$: we have $t_b(f_\gamma=\pi^2)=0$, and $t_b'(f_\gamma=\pi^2)=0$. Moreover (\ref{equa:tbuck}) also shares the large $f_\gamma$ limit of (\ref{equa:buckling-curve}) and (\ref{equa:buckling-curve-approx-1}): $t_b =  - 4 \pi^2 + f_\gamma + \ldots$. We see in Figure~\ref{fig:bucklingcurve} that the curves defined by (\ref{equa:buckling-curve-approx-1}) and (\ref{equa:tbuck}) are in fact hard to distinguish.

\section{Experimental verification} \label{section:exp}
%
%
%
%

%
In order to verify experimentally and quantitatively the mechanics of the windlass, we place a drop on a fiber and test whether the fibre coils in the drop. The fibre is made of BASF Thermoplastic PolyUrethan (TPU) that is melt-spinned. This process involves melting down the TPU, then applying a large extension rate to the liquid filament while it cools down rapidly in the ambient air. It results in reproducible, meter-long micronic fibres (1-20 $\mu$m in radius) with portions away from the edges having small perturbations in radius (typically <5\% every 1000 radii). Calipers are used to further manipulate the samples. Clamping is achieved with cured Loctite\textsuperscript{\textregistered} glue. The system size is measured optically with a Leica macroscope (VZ85RC) mounted on a micro-step motor and a 3 megapixels Leica DFC-295 camera ($400\times$ magnification, 334nm/pixel picture resolution) with a Phlox 50x50 mm backlight, at 60000 lux or alternatively an optical fibre with LED lamp (Moritex MHF-M1002) with circular polarizer. The fibre radius is then extracted by image analysis, using imageJ (http://imagej.nih.gov/ij/).
For the droplet, we select silicone oil Rhodorsil 47V1000, the figure 1000 referring to its viscosity compared to water. High viscosity was chosen in order to be able to deposit drops on the fibre by brushing, and for its slow evaporation properties. Using the condition that meniscus forces have to support the weight of the droplet and be strong enough to buckle the beam, we find that to be bendable, a TPU fibre must be below 7.2 $\mu$m in radius, with $\text{E}_{\text{TPU}}=17 \pm 3 \text{ MPa, } \gamma_{\text{silicone-oil/air}} = 21.1 \text{mN/m}, \alpha_{Y, \text{ silicone-oil/TPU}} = 27 \pm 5 \text{ degrees}$ and $\rho_{\text{silicone-oil}} = 960 \text{ kg} / \text{m}^3$ the silicone oil density.

\begin{figure}[h]
\centering
\includegraphics[width=.5\columnwidth]{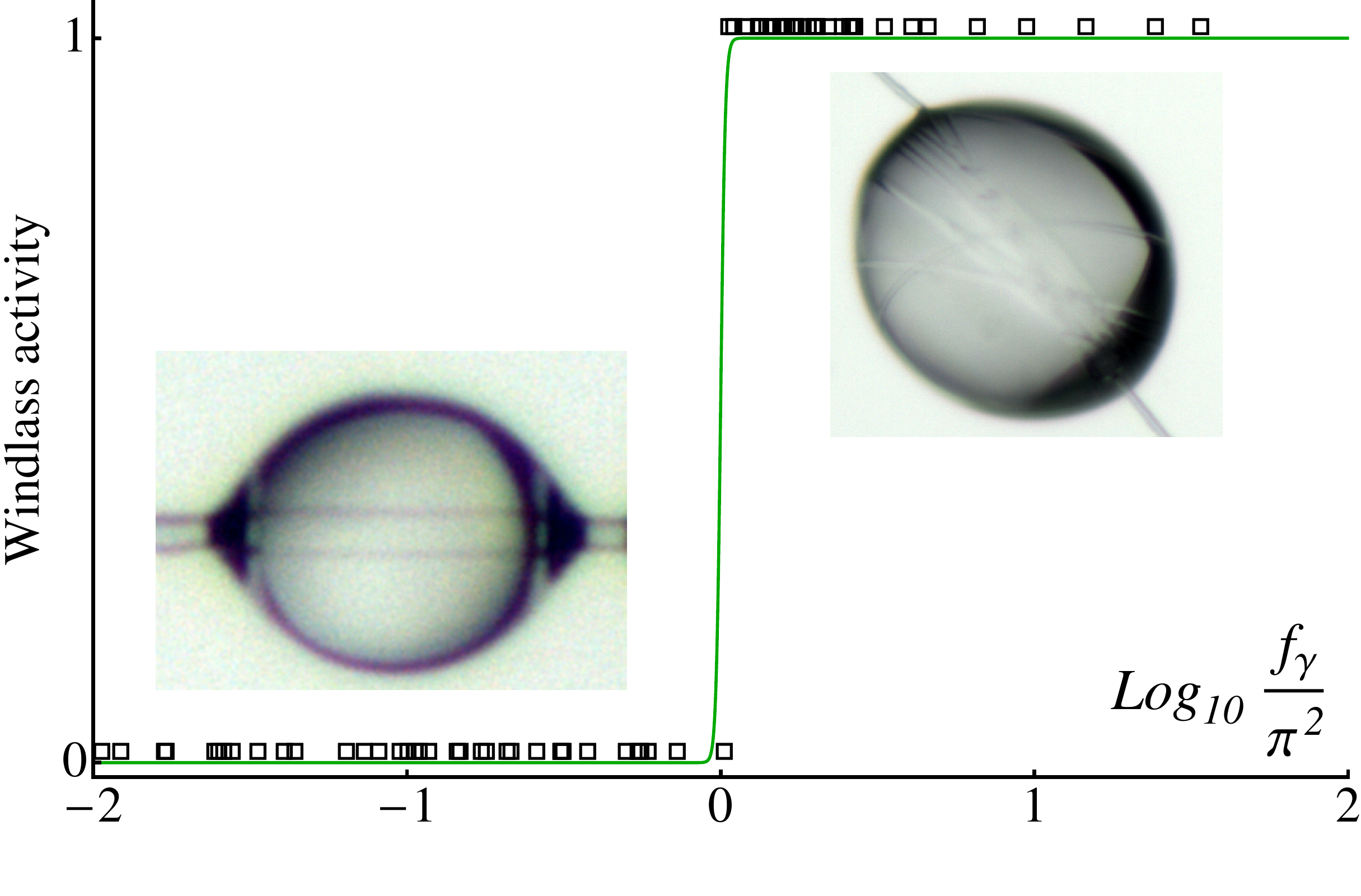}
\vspace{-1em}
\caption{Experimental verification of the windlass activation as function of the parameter $f_{\gamma}$. The windlass mechanism is active as soon as the meniscus force $f_\gamma$ is greater than $\pi^2$.}
\label{fig:windlass-activation}
\end{figure}

We deposit a drop on a fibre with a known radius and we slowly bring the caliper forks closer to impose compression on the fibre. If the drop is able to coil the fibre, the macroscopic consequences are easily visible to the naked eye. If at first try the drop does not coil, the sample is tested again with a random compression and a slight shake to overcome any possible metastability. The couple drop/fiber is given an activity index of 1 if coiling is achieved and 0 otherwise (see figure \ref{fig:windlass-activation}). The experimental results show that coiling is present whenever $f_\gamma > \pi^2$, with a error of margin of 3\%, consistent with equation (\ref{equa:buckling-curve}).

\section{Nonlinear post-buckling computations} \label{sec:post-buckling}
%
%
%
%
\begin{figure}[h]
\centering
\includegraphics[width=.47\columnwidth]{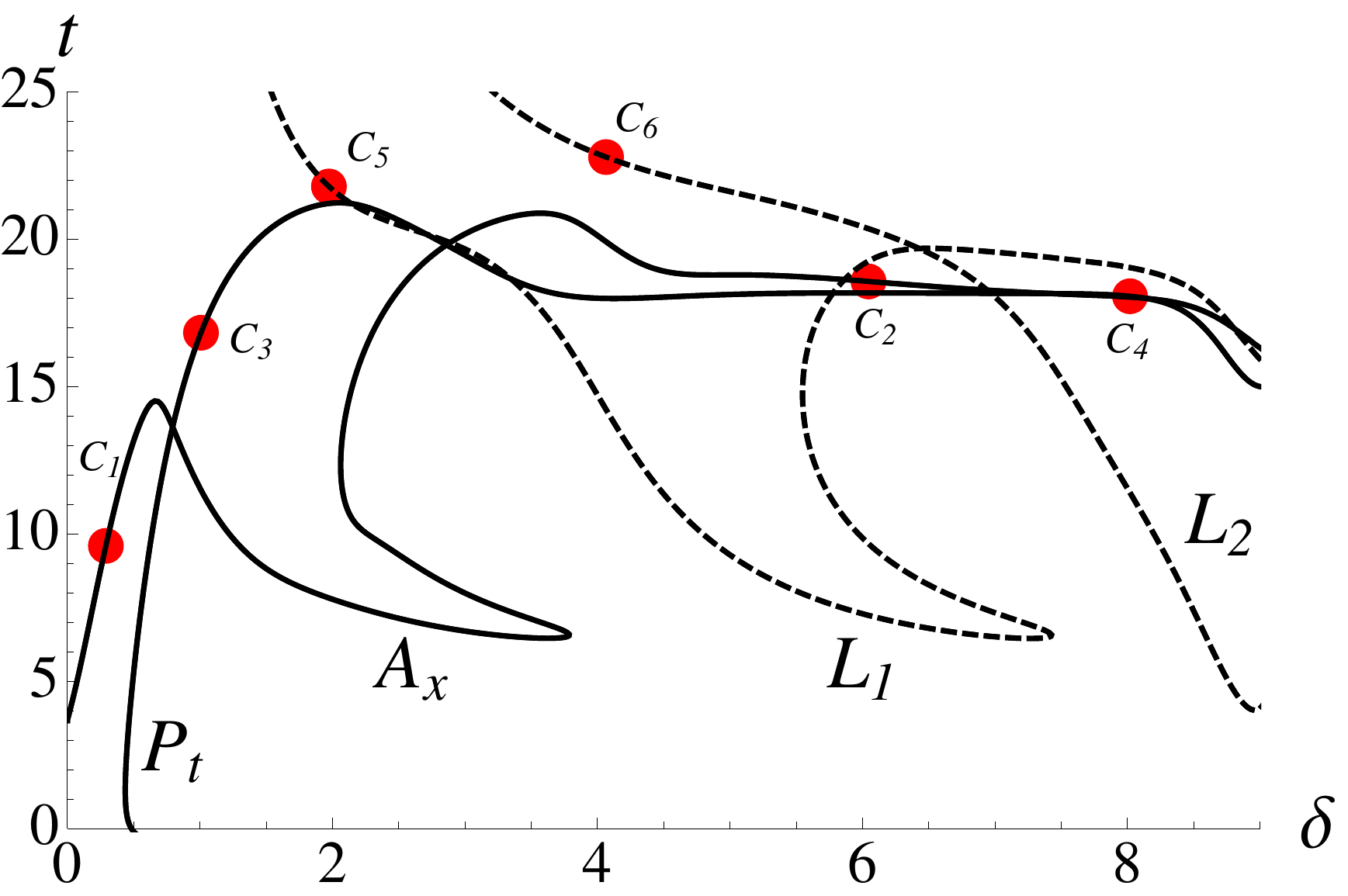}
\includegraphics[width=.485\columnwidth]{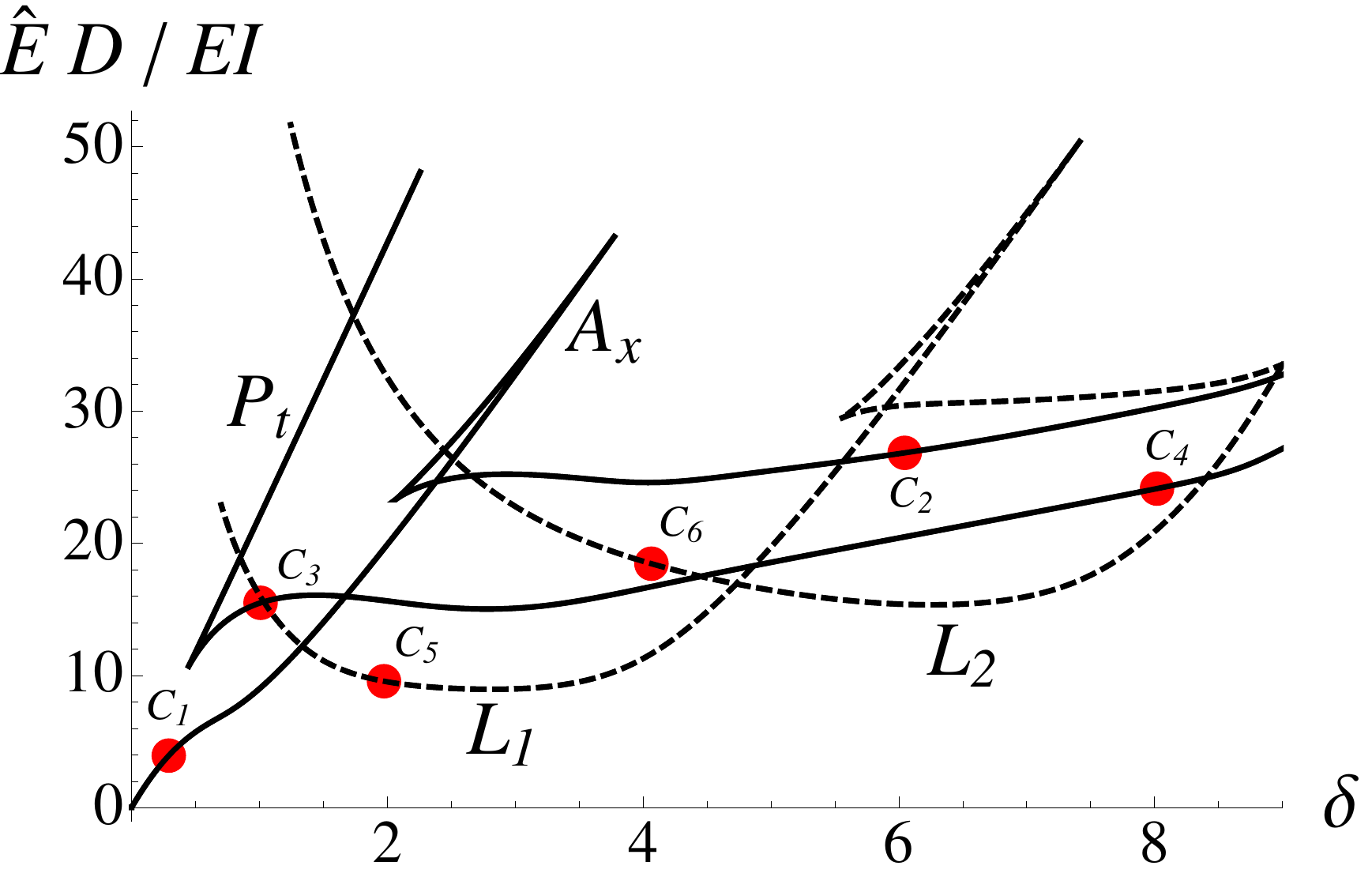}
\caption{Post-buckling paths: (Left) Force-displacement curves for $f_\gamma=20$, where $t$ is the applied tension and $\delta$ the end-shortening, and (Right) Energy $\hat{E}$ as function of $\delta$. Solid lines represents both $Ax$ configurations, where the beam shape is symmetric with respect to the axis joining the center of the disk $(x_C,y_C)$ and the beam midpoint $(x(\ell/2),y(\ell/2))$, and $Pt$ configurations, where the beam shape is symmetric with respect to the beam midpoint $Pt$. Dashed lines correspond to $L_1$ and $L_2$ configurations.}
\label{fig:bif-path-fg-20}
\end{figure}
\begin{figure}[h]
\centering
\includegraphics[width=.48\columnwidth]{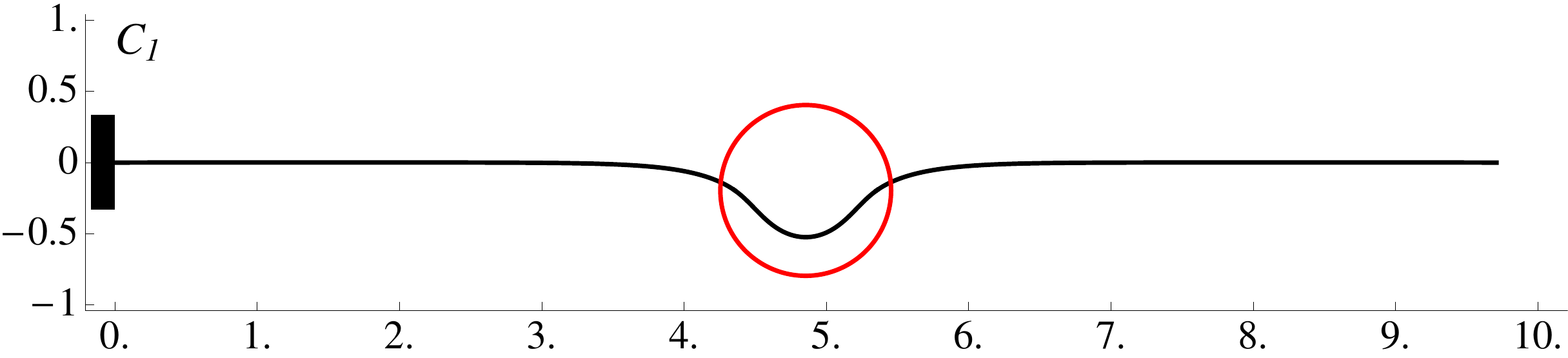} ~ \includegraphics[width=.48\columnwidth]{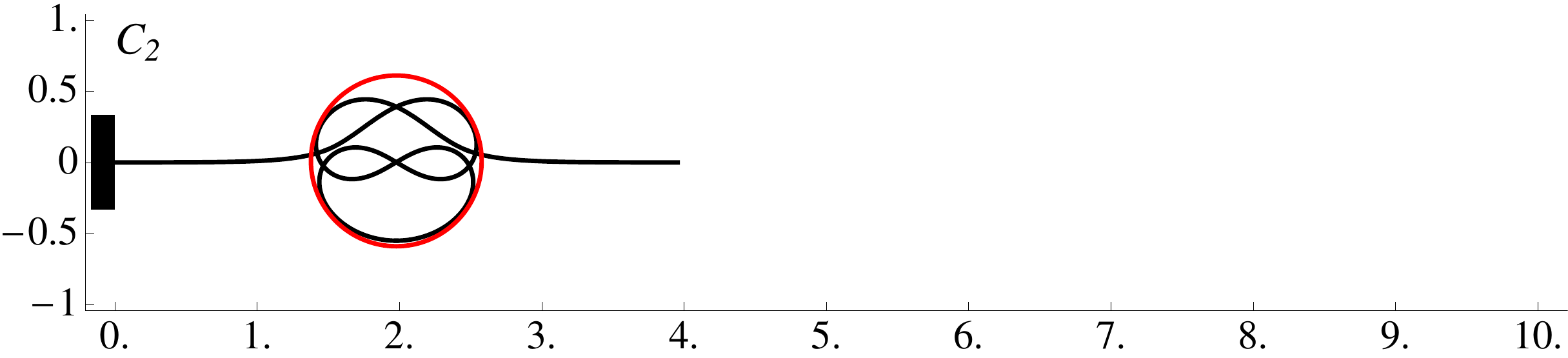}\\
\includegraphics[width=.48\columnwidth]{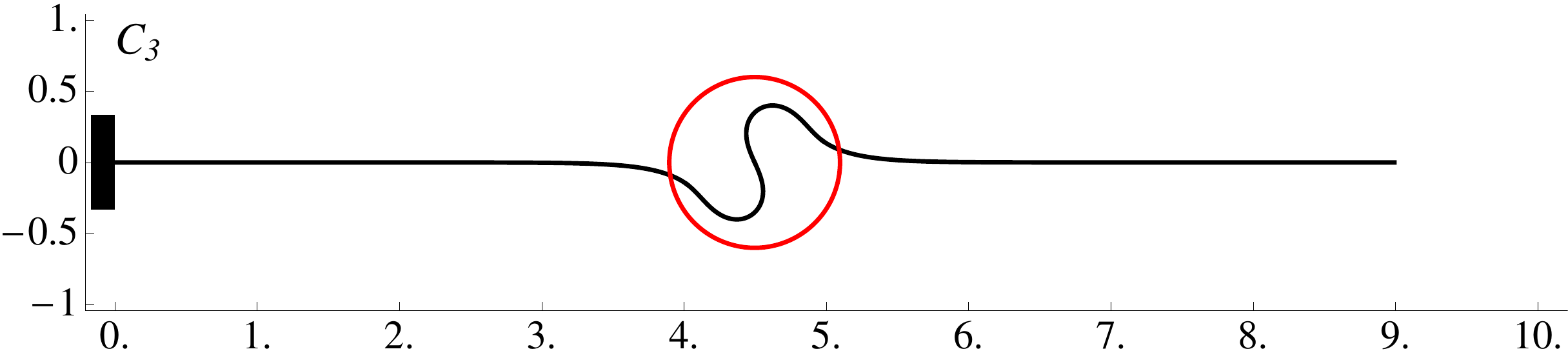} ~ \includegraphics[width=.48\columnwidth]{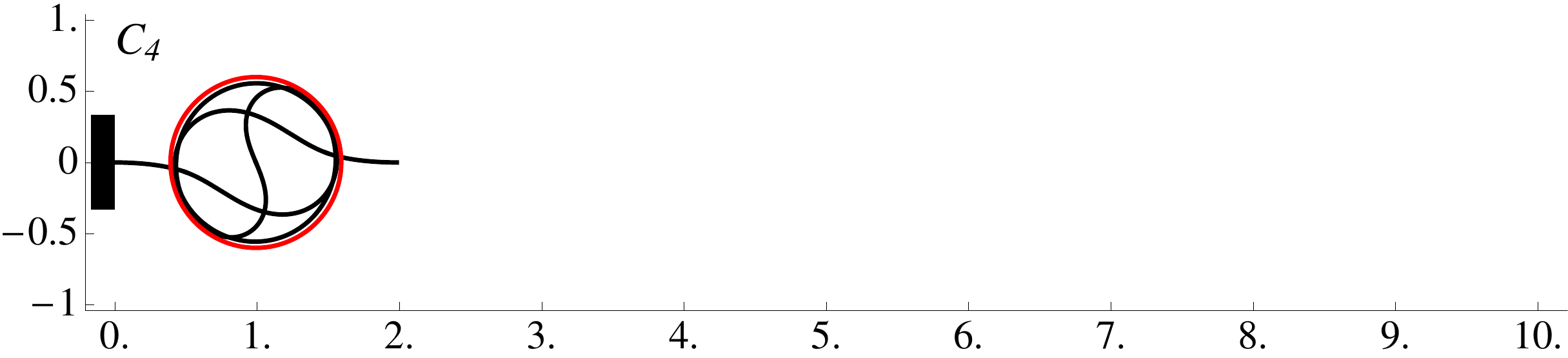}\\
\includegraphics[width=.48\columnwidth]{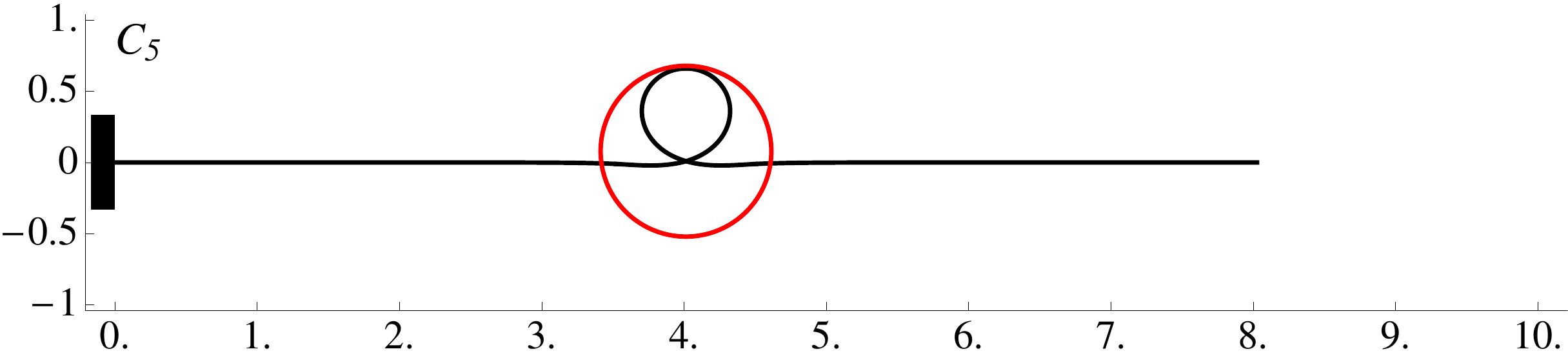} ~ \includegraphics[width=.48\columnwidth]{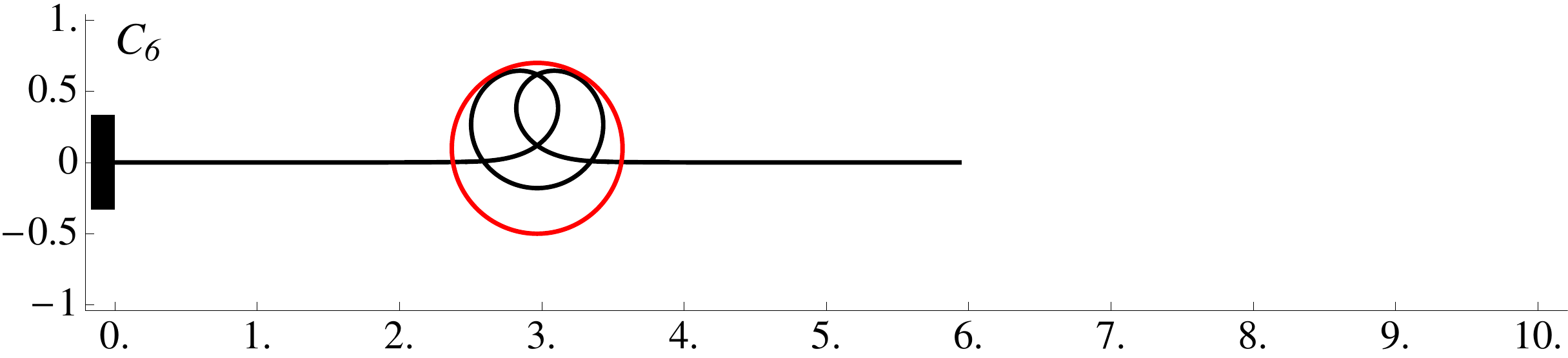}\\
\caption{Post-buckling configurations. Top row shows $Ax$ configurations, symmetric with respect to the axis joining the center of the disk $(x_C,y_C)$ and the beam midpoint $(x(\ell/2),y(\ell/2))$. Middle row shows $Pt$ configurations, symmetric with respect to the beam midpoint $Pt$. Bottom row represents $L_1$ and $L_2$ configurations. The circles have center $(x_C,y_C)$ and radius $(1+\rho)/2$.}
\label{fig:shapes-fg-20}
\end{figure}
%
%

%
%
%
%
\subsection*{Equilibrium paths} \label{section:equilibrium-solutions}
%
%
%
%
We now analyze the post-buckling regime by numerically solving the non-linear system of equilibrium equations. 
We use a shooting method to solve the boundary-value problem (\ref{sys:equilibre-adim})-(\ref{equa:sauts-force}) and a pseudo-arc-length continuation algorithm to follow the solution as parameters are varied, both of these routines being implemented in Mathematica. For large $f_\gamma$ values, typically $f_\gamma > 15$, numerical difficulties arise and we thankfully switch to the AUTO package~\cite{Doedel1991Numerical-Analysis}.
We fix $\ell=10$, 
$v_0=0.02$,   
$\rho=0.2$, 
and we compute force-extension bifurcation diagrams for several values of $f_\gamma$.
We show in Figure~\ref{fig:bif-path-fg-20} such a diagram for $f_\gamma=20$ where the tension $t$ is plotted as function of the end-shortening $\delta = \ell - x(\ell)$. 
The diagram comprises four different equilibrium paths: $(i)$ path $Ax$ where configurations are symmetric with respect to the axis joining the center of the disk $(x_C,y_C)$ and the beam midpoint $(x(\ell/2),y(\ell/2))$, $(ii)$ path $Pt$ where configurations are symmetric with respect to the beam midpoint, $(iii)$ path $L_1$ where configurations are looping once inside the disk, and $(iv)$ path $L_2$ where configurations are looping twice inside the disk. Few of these configurations are shown in Figure~\ref{fig:shapes-fg-20}. 
In a typical experiment the system is first completely straight, held by a large tension $t$. This situation corresponds to a point on the vertical axis of Figure~\ref{fig:bif-path-fg-20}, above the buckling threshold. As $t$ is decreased the systems reaches the start of the $Ax$ path, and the beam buckles. 
For $f_\gamma=20$, the numerically found value of the buckling tension $t\simeq 3.66$ is to be compared to $t_b(20) \simeq 3.76$ given by (\ref{equa:tbuck}), and $t \simeq 3.73$ given by (\ref{equa:buckling-curve}).
As the system branches on the $Ax$ path, tension goes up again --- we have a subcritical bifurcation.
The slope of the $Ax$ path is calculated analytically in \ref{section:weakly-non-lin} and is plotted in Figure~\ref{fig:fnl} for comparison.
As the beam enters deeper in the post-buckling regime, bending localizes inside the disk and the tails remain approximatively straight. The path eventually reaches a plateau, see formula (\ref{equa:t_plateau}), where the beam coils in a circular way inside the disk and the bending energy in the beam can then be approximated by $(1/2) \, EI/R^2 \, [L-X(L)]$.
We also plot the $t>0$ part of the path $Pt$. This path also reaches the same plateau as the beam coils in the same circular way inside the disk.
In addition we plot paths along which the beam adopts configurations with one (path $L_1$) or two (path $L_2$) loops.
The relevance of these paths could be questioned for two reasons: $(i)$ configuration on path $L_1$ do not have the same topology as far as twist is considered: a full turn of twist would be necessary to connect configurations on path $Ax$ or $Pt$ with configurations on path $L_1$, see \cite{heijden+al:2001,Goss2005}, and $(ii)$ these paths are not connected to the vertical axis $\delta=0$.
We plot in Figure~\ref{fig:bif-path-fg-20}-Right the energy $\hat{E}=E_\kappa+E_w+E_\gamma-P\, \gsv \, L + F_\gamma (\Delta+D)$ as a function of the end-shortening $\delta$ and we see that, for some range of the end-shortening $\delta$, configurations on paths $L_1$ or $L_2$ have a lower energy than configurations on paths $Ax$ or $Pt$.
These remarks call for a stability analysis of the equilibrium configurations, as well as a study of configurations deformed in $3D$, where twist, link, and writhe would be computed \cite{Elettro2015b}.

\subsection*{Approximate analytical model for the plateau regime} \label{sec:approx}
\begin{figure}
\centering
\includegraphics[width=.45\columnwidth]{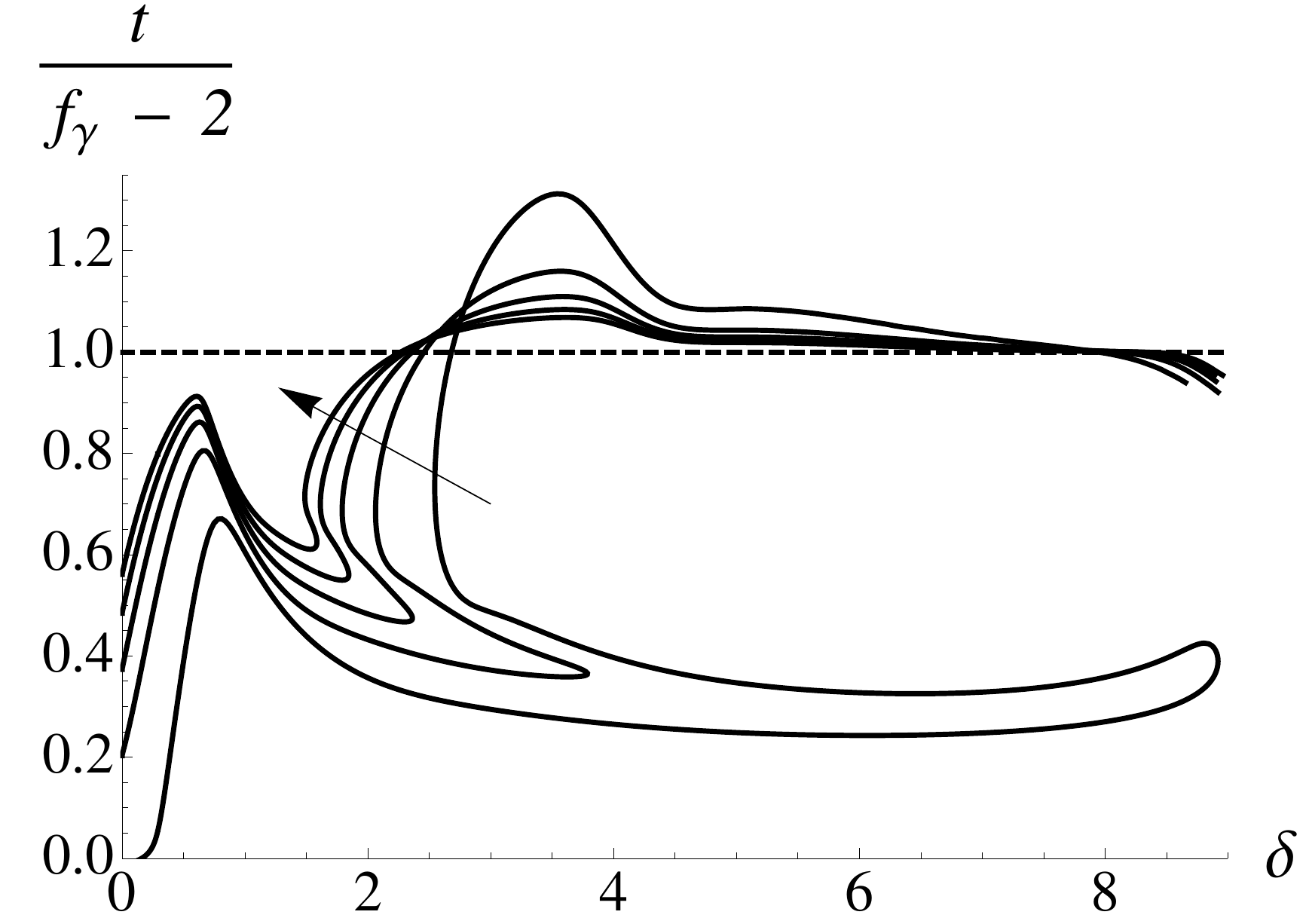} ~ ~
\includegraphics[width=.45\columnwidth]{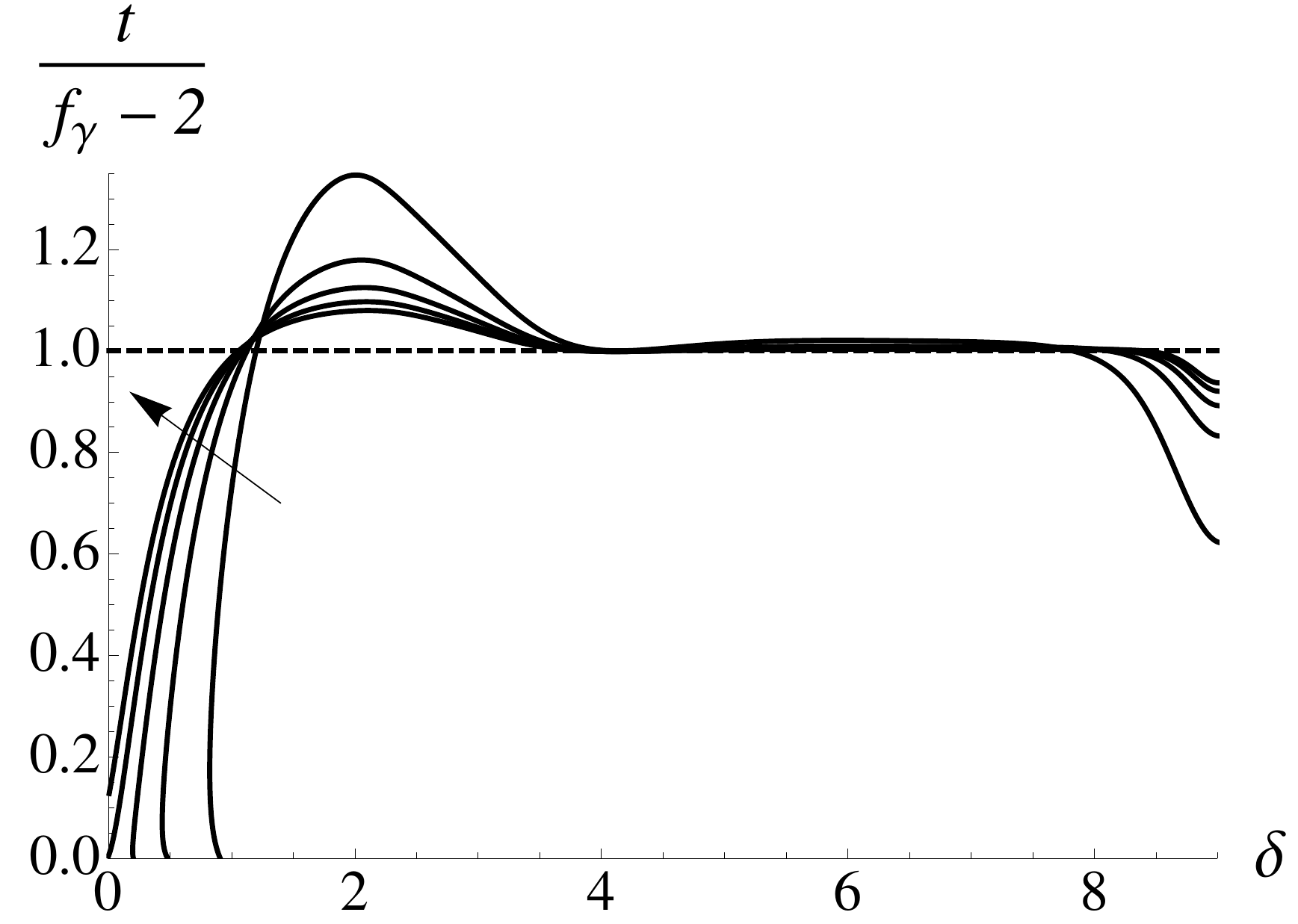}
\caption{
Equilibrium paths for $f_\gamma=10, 20, \ldots, 50$ with the vertical axis rescaled according to Equation (\ref{equa:t_plateau}).
(Left) Axis-symmetric ($Ax$) configurations  and (Right) Point-symmetric  ($Pt$) configurations. Direction of increasing $f_\gamma$ is indicated with the arrows.}
\label{fig:plateau}
\end{figure}

%
%
%
%
As explained in \cite{Elettro2015In-drop-capillary}, in the regime where the end-shortening $\delta = \ell - x(\ell)$ is large, that is when several coils of the beam are present in the disk, the external tension $t$ reaches a plateau and no longer varies as more coils are added.
The plateau value of the tension is calculated by a balance of energy as a beam length $\Delta S$ enters the disk.
The work done by the tension $T$ is $- T \Delta S$, the work done by the meniscus force $F_\gamma$ is $+ F_\gamma \, \Delta S$, and the energy spent to bend the beam in coils is $-(1/2) \, (EI/R^2)  \, \Delta S$. The sum of these energies is zero on the plateau, which yields
\begin{equation}
t = f_\gamma - 2
\label{equa:t_plateau}
\end{equation}
in dimensionless quantities.
In Figure~\ref{fig:plateau} we plot  $Ax$ and $Pt$ equilibrium paths for $f_\gamma=10, 20, \ldots, 50$ with the vertical axis rescaled according to (\ref{equa:t_plateau}). The collapse of the curves for $\delta \gtrsim 4$ confirms relation (\ref{equa:t_plateau}).
\section{Conclusion}
%
%
%
This paper presents a first venture into the complex equilibria adopted by a fiber buckled, coiled, and packed by a droplet. Using a simple 2D model and numerical continuation techniques, we have uncovered several equilibrium paths characterized by different symmetries (point-symmetric, axis-symmetric, single or double-looped) and provided clues for the bifurcations between these different states. Analytical predictions for the buckling threshold, as well as the asymptotic behavior (plateau regime) for the deep post-buckling regime have been derived. The agreement between experiments and theory for the windlass activation threshold is certainly promising, and calls for an extension of the present model to 3D (including twist and writhe), and a deeper comparison between the experimentally observed ordered and disordered packing modes and the theoretical prediction. Finally the deformation of the drop on its own, considered rigid throughout this study, and its interplay with the shape adopted by the fiber also deserves a dedicated investigation.
%


\section*{Acknowledgments}
The present work was supported by ANR grant ANR-09-JCJC-0022-01, ANR-14-CE07-0023-01, and ANR-13-JS09-0009. Financial support from `La Ville de Paris - Programme \'Emergence' is also gratefully acknowledged, along with travel support from the Royal Society, through the International Exchanges Scheme (grant IE130506) and support from the CNRS, through a PEPS-PTI grant.
We thank Christine Rollard (MNHN) for helpful discussions.
%
%
%
%
%
%
%
%
%
%
%
%
%
\appendix
\section{Variational derivation of the equilibrium equations} \label{section:varia}
%
%
%
%
%
%
%
%
%
%
To prevent the beam from exiting the disk elsewhere than at the meniscus points, we use a (soft wall) barrier potential \cite{ManningBulman:2005}
\begin{equation}
V(X,Y) = \frac{V_0}{1 + \rho - (1/R) \sqrt{(X-X_C)^2 + (Y-Y_C)^2}   }
\end{equation}
where the disk has center $(X_C,Y_C)$ and radius $R$. The small dimensionless parameter $\rho$ is introduced to avoid the potential to diverge at the meniscus points $A$ and $B$, where the rod enters and exits the disk.
The internal energy of the system comprises the bending energy $E_b$ of the rod, the barrier energy $E_w$ of the circle, and surface energy $E_\gamma$:
\begin{subequations}
\begin{align}
E_\kappa &=
 \frac{1}{2}EI \int_0^{S_A}  \kappa_1^2 \, \d S +
 \frac{1}{2}EI \int_{S_A}^{S_B}  \kappa_2^2 \, \d S +
 \frac{1}{2}EI \int_{S_B}^{L} \kappa_3^2 \, \d S \\
E_w &  = \int_{S_A}^{S_B}  V(X(S),Y(S),X_C,Y_C) \, \d S    \\
E_\gamma &= P \, \gsv \, S_A + P \, \gsl \, (S_B-S_A) + P \, \gsv \, (L - S_B)
\end{align}
\end{subequations}
where  $P=2\pi \, h$ is the perimeter of the cross-section of the rod, and $V_0$ has the dimension of an energy per unit length.
The curvatures $\kappa_i(S)$ are defined in each region of the rod.
We add the work done by the external load $T \, \bm{e_x}$ and obtain the potential energy of the system:
\begin{equation}
E_\kappa + E_w + E_\gamma - T \, X(L)
\end{equation}
We minimize this energy under the following constraints
\begin{subequations}
\begin{align}
\frac{S_A + S_B}{2} &= \frac{L}{2} \label{equa:app_sasb} \\
\left[ X(S_A) - X_C \right]^2 + [ Y(S_A) - Y_C]^2 &= R^2 \label{equa:A4b} \\
[ X(S_B) - X_C ]^2 + [ Y(S_B) - Y_C]^2 &= R^2 \label{equa:A4c}
\end{align}
\end{subequations}
Equation (\ref{equa:app_sasb}) imposes that the capturing disk is centered on the mid-point of the rod. We introduce $\Sigma$ such that $S_A = \sa$ and $S_B=\sb$. The rod has then $2 \, \Sigma$ of its arc-length spent in the disk. As the variables $X(S)$, $Y(S)$, $\kappa(S)$ and $\theta(S)$ all appear in the formulation, we have to consider the continuous constraints relating them:
\begin{equation}
X'(S)= \cos \theta(S) , \; Y'(S)= \sin \theta(S) , \; \theta'(S) = \kappa(S)
\label{equa:app-continuous-constraints}
\end{equation}
We consequently write the Lagrangian:
\begin{subequations}
\begin{align}
{\cal L}(X,&Y,\theta,\kappa_1,\kappa_2,\kappa_3,X_C,Y_C,\Sigma) = -T X(L) + \nonumber \\
& \int_0^{\sa} \left( \frac{EI}{2} \kappa_1^2   + P \, \gsv + \nu_1(S) \left[ X'-\cos \theta \right] +  \mu_1(S) \left[ Y'-\sin \theta \right] + \eta_1(S) \left[ \theta'-\kappa_1 \right] \right) \d S + \nonumber \\
& \int_{\sa}^{\sb} \left( \frac{EI}{2} \kappa_2^2   + P \, \gsl + \nu_2(S) \left[ X'-\cos \theta \right] +  \mu_2(S) \left[ Y'-\sin \theta \right] + \eta_2(S) \left[ \theta'-\kappa_2 \right] + V(X,Y,X_C,Y_C)\right) \d S + \nonumber\\
& \int_{\sb}^{L} \left(\frac{EI}{2} \kappa_3^2   + P \, \gsv + \nu_3(S) \left[ X'-\cos \theta \right] +  \mu_3(S) \left[ Y'-\sin \theta \right] + \eta_3(S) \left[ \theta'-\kappa_3 \right] \right) \d S  + \nonumber \\
& \frac{\Lambda_A}{2R} \left( \left[ X\left(\sa \right) - X_C \right]^2 +
 \left[ Y\left(\sa \right) - Y_C \right]^2 -R^2\right) + \nonumber \\
&\frac{\Lambda_B}{2R} \left( \left[ X\left(\sb \right) - X_C \right]^2 +
\left[ Y\left(\sb \right) - Y_C \right]^2 -R^2\right)
\end{align}
\end{subequations}
The rod is clamped as both extremities, boundary conditions reads:
\begin{equation}
X(0) = 0 , \; 
Y(0) = 0 , \; 
\theta(0) =0 , \;
Y(L) = 0 , \; 
\theta(L) =0 
\label{equa:app_BC}
\end{equation}
\subsection*{First variation}
%
%
%
%
%
We note $\bm{U} = (X,Y,\theta,\kappa_1,\kappa_2,\kappa_3,X_C,Y_C,\Sigma)$ and we consider the conditions for the state $\bm{U}_e$ to minimize the energy $E$. Calculus of variations shows that a necessary condition is 
\begin{equation}
{\cal L}'(\bm{U}_e)\bar{\bm{U}} = \left. \frac{\d}{\d \epsilon} {\cal L}(\bm{U}_e + \epsilon \, \bar{\bm{U}}) \right|_{\epsilon=0}= 0
\label{equa:app_1st-varia}
\end{equation}
where $\bar{\bm{U}} = (\bar X, \bar Y, \bar \theta, \bar \kappa_1, \bar \kappa_2, \bar \kappa_3, \bar X_C, \bar Y_C, \bar \Sigma)$. The bar ~ $\bar{}$ ~ sign represents a small perturbation of the variable. Moreover boundary conditions (\ref{equa:app_BC}) imply that 
\begin{equation}
\bar X(0) = 0  , \; \bar Y(0) = 0 , \;  \bar \theta(0) =0 ,  \;  \bar Y(L) = 0 ,  \; \bar \theta(L) =0
\label{equa:app-BC-bar}
\end{equation}
Noting that
$\int_0^{A+\epsilon \bar{A}} f(x) dx = \int_0^{A} f(x) dx + \epsilon \bar{A} f(A) + O(\epsilon^2)$
we evaluate the first variation (\ref{equa:app_1st-varia})
\begin{subequations}
\label{sys:app-1st-varia}
\begin{align}
{\cal L}'(U_e)(\bar{U})  &= -T \bar X(L) - 2 F_\gamma \, \bar \Sigma + V_A \, \bar \Sigma + V_B \, \bar \Sigma + \nonumber \\
&\int_0^{\sa} \left( EI \, \bar \kappa_1 \, \kappa_1   + \nu_1(S) \left[ \bar X' + \bar \theta \sin \theta \right] +  \mu_1(S) \left[ \bar Y' - \bar \theta \cos \theta \right] + \eta_1(S) \left[ \bar \theta'- \bar \kappa_1 \right] \right) \d S + \nonumber \\
& \int_{\sa}^{\sb} \left( EI \, \bar \kappa_2 \, \kappa_2   +  \nu_2(S) \left[ \bar X' + \bar \theta \sin \theta \right] +  \mu_2(S) \left[\bar  Y' - \bar \theta \cos \theta \right] + \eta_2(S) \left[\bar \theta'- \bar \kappa_2 \right] + \bar V\right) \d S \nonumber\\
& \int_{\sb}^{L} \left(EI \, \bar \kappa_3\,  \kappa_3  + \nu_3(S) \left[\bar X' + \bar \theta \sin \theta \right] +  \mu_3(S) \left[ \bar Y'- \bar \theta \cos \theta \right] + \eta_3(S) \left[\bar \theta'- \bar \kappa_3 \right] \right) \d S  + \nonumber \\
& \frac{\Lambda_A}{R} 
\left[ X\left(\sa \right) - X_C \right] \left[ \bar X\left(\sa \right) - \bar\Sigma \,X'\left(\sa \right) - \bar X_C \right]  + \nonumber \\
&\frac{\Lambda_A}{R} 
\left[ Y\left(\sa \right) - Y_C \right] \left[ \bar Y\left(\sa \right) - \bar\Sigma \,Y'\left(\sa \right) - \bar Y_C \right]  + \nonumber \\
& \frac{\Lambda_B}{R} 
\left[ X\left(\sb \right) - X_C \right] \left[ \bar X\left(\sb \right) + \bar\Sigma \,X'\left(\sb \right) - \bar X_C \right]  + \nonumber \\
&\frac{\Lambda_B}{R} 
 \left[ Y\left(\sb \right) - Y_C \right] \left[ \bar Y\left(\sb \right) + \bar \Sigma \,Y'\left(\sb \right) - \bar Y_C \right]  
\end{align}
\end{subequations}
where $F_\gamma = P \, (\gsv - \gsl)$, $V_A=V(X(S_A),Y(S_A))$, $V_B=V(X(S_B),Y(S_B))$, and 
$\bar V = (\partial V / \partial X) \,  \bar X + (\partial V / \partial Y) \,  \bar Y + (\partial V / \partial X_C) \,  \bar X_C + (\partial V / \partial Y_C) \,  \bar Y_C$. Note also that we have used (\ref{equa:app-continuous-constraints}) at $S=L/2\pm\Sigma$ to eliminate several terms.
Requiring (\ref{sys:app-1st-varia}) to vanish for all $\bar \kappa_i$, $i=1,2,3$, we obtain
\begin{equation}
EI \kappa_i(S) = \eta_i(S), \quad i=1,2,3
\end{equation}
and hence identify the Lagrange multipliers $\eta_i(S)$ with the bending moment $M(S)$ in the beam.
Requiring (\ref{sys:app-1st-varia}) to vanish for all $\bar \theta$ yields, after integration by parts: 
\begin{subequations}
\label{sys:app-theta}
\begin{align}
  \left[ \eta_1(S) \, \bar \theta \, \right]_0^{\sa} + \int_0^{\sa} \left( \nu_1(S)  \sin \theta -  \mu_1(S)  \cos \theta - \eta'_1(S)  \right) \bar \theta \, \d S \; + \nonumber \\
  \left[ \eta_2(S) \, \bar \theta \, \right]_{\sa}^{\sb}  + \int_{\sa}^{\sb} \left(   \nu_2(S) \sin \theta  -  \mu_2(S) \cos \theta - \eta'_2(S) \right) \bar \theta \, \d S \; &+ \nonumber\\
  \left[ \eta_3(S) \, \bar \theta \, \right]_{\sb}^{L}  + \int_{\sb}^{L} \left(   \nu_3(S) \sin \theta  -  \mu_3(S) \cos \theta - \eta'_3(S) \right) \bar \theta \, \d S \; & = 0
\end{align}
\end{subequations}
Due to the boundary conditions (\ref{equa:app-BC-bar}), part of the boundary terms vanish. Nevertheless, arbitrariness of $\bar \theta$ at $S=L/2 \pm \Sigma$ implies that  $\eta_1\left(\sa \right)= \eta_2\left(\sa \right)$ and $\eta_2\left(\sb \right)= \eta_3\left(\sb \right)$: the bending moment is continuous at the entry and the exit of the disk.
Moreover, from the requirement that (\ref{sys:app-theta}) vanishes for all $\bar \theta(S)$, we obtain the equations for the equilibrium of the bending moment
\begin{equation}
\eta'_i(S) =M'(S) = \nu_i(S)  \sin \theta -  \mu_i(S)  \cos \theta , \quad i=1,2,3
\end{equation}
Requiring (\ref{sys:app-1st-varia}) to vanish for all $\bar X$ yields, after integration by parts: 
\begin{subequations}
\label{sys:app-x}
\begin{align}
\frac{\Lambda_A}{R} \left[ X\left(\sa \right) - X_C \right] \, \bar X\left(\sa \right) &+  \left[ \nu_1(S) \, \bar X \, \right]_0^{\sa} - \int_0^{\sa}  \nu'_1(S) \, \bar X \, \d S  \; - \nonumber\\
T \bar X(L)  & +  \left[ \nu_2(S) \, \bar X \, \right]_{\sa}^{\sb}  + \int_{\sa}^{\sb} 
   \left[  \frac{\partial V}{\partial X} - \nu'_2(S) \right] \bar X \, \d S \; + \nonumber\\
\frac{\Lambda_B}{R} \left[ X\left(\sb \right) - X_C \right] \, & \bar X\left(\sb \right) +   \left[ \nu_3(S) \, \bar X \, \right]_{\sb}^{L}  - \int_{\sb}^{L}  \nu'_3(S) \,  \bar X \, \d S \;  =0 
\end{align}
\end{subequations}
Boundary conditions (\ref{equa:app-BC-bar}) cancel part of the boundary terms, but arbitrariness of $\bar X$ at $S=L/2 \pm \Sigma$ implies 
\begin{subequations}
\label{sys:app-saut-forces-x}
\begin{align}
\nu_2\left(\sa \right) -\nu_1\left(\sa \right) &= \frac{\Lambda_A}{R} \left[ X\left(\sa \right) - X_C \right] 
\label{equa:app-force-jumpAx} \\
\nu_3\left(\sb \right) -\nu_2\left(\sb \right) &= \frac{\Lambda_B}{R} \left[ X\left(\sb \right) - X_C \right]
\label{equa:app-force-jumpBx}
\end{align}
\end{subequations}
while arbitrariness of $\bar X$ at $S=L$ implies 
\begin{equation}
\nu_3(L) = T \label{equa:app-force-finale}
\end{equation}
Equation (\ref{equa:app-force-finale}) enable us to identify $\nu_3$ and therefore $\nu_1$ and $\nu_2$ with the $x$-component, $N_x$, of the resultant force in the beam. By extension the $\mu_i$ are identified to the $y$-component, $N_y$, of this force.
Equations (\ref{sys:app-saut-forces-x}) are then seen as jumps in the $x$-component of the internal force due to the external force coming from the disk.
Moreover, from the requirement that (\ref{sys:app-x}) vanishes for all $\bar X(S)$, we obtain the equilibrium equations for the $x$-component of  the resultant force in the beam
\begin{equation}
\nu'_1(S) = 0 \, , \quad \nu'_2(S) =  \frac{\partial V}{\partial X} \, , \quad \nu'_3(S) = 0
\label{equa:app-equil-forcesX}
\end{equation}
The same procedure for the variable $Y$ yields
\begin{subequations}
\label{sys:app-saut-forces-y}
\begin{align}
\mu_2\left(\sa \right) -\mu_1\left(\sa \right) &= \frac{\Lambda_A}{R} \left[ Y\left(\sa \right) - Y_C \right]
\label{equa:app-force-jumpAy}  \\
\mu_3\left(\sb \right) -\mu_2\left(\sb \right) &= \frac{\Lambda_B}{R} \left[ Y\left(\sb \right) - Y_C \right]
\label{equa:app-force-jumpBy}
\end{align}
\end{subequations}
and
\begin{equation}
\mu'_1(S) = 0 \, , \quad \mu'_2(S) =  \frac{\partial V}{\partial Y} \, , \quad \mu'_3(S) = 0
\label{equa:app-equil-forcesY}
\end{equation}
Equations (\ref{sys:app-saut-forces-y}) are then seen as jumps in the $y$-component of the internal force due to the external force coming from the disk, and equations (\ref{equa:app-equil-forcesY}) are the equilibrium equations for the $y$-component of  the resultant force in the beam.
Requiring (\ref{sys:app-1st-varia}) to vanish for all $\bar X_C$ yields
\begin{equation}
-\frac{\Lambda_A}{R} \left[ X\left(\sa \right) - X_C \right] 
-\frac{\Lambda_B}{R} \left[ X\left(\sb \right) - X_C \right] 
+\int_{\sa}^{\sb}   \frac{\partial V}{\partial X_C} \, \d S = 0
\end{equation}
We use the identity $\frac{\partial V}{\partial X_C} = -\frac{\partial V}{\partial X}$ and (\ref{equa:app-equil-forcesX}) to obtain
\begin{equation}
\nu_2\left(\sb \right) -\nu_2\left(\sa \right)   = - \frac{\Lambda_A}{R} \left[ X\left(\sa \right) - X_C \right]  - \frac{\Lambda_B}{R} \left[ X\left(\sb \right) - X_C \right] 
\label{equa:app-XC}
\end{equation}
The same procedure for the variable $Y_C$ yields
\begin{equation}
\mu_2\left(\sb \right) -\mu_2\left(\sa \right)   = - \frac{\Lambda_A}{R} \left[ Y\left(\sa \right) - Y_C \right]  - \frac{\Lambda_B}{R} \left[ Y\left(\sb \right) - Y_C \right] 
\label{equa:app-YC}
\end{equation}
Considering (\ref{equa:app-XC}), (\ref{equa:app-YC}), (\ref{sys:app-saut-forces-x}), and  (\ref{sys:app-saut-forces-y}) together yields
\begin{align}
\nu_1\left(\sa \right) &= \nu_3\left(\sb \right) \\
\mu_1\left(\sa \right) &= \mu_3\left(\sb \right) \label{eq:A24}
\end{align}
which means that the internal force in the beam at the entrance of the disk is equal to the internal force at the exit of the disk. We therefore have that the total external force applied on the beam by the disk is zero. 
Finally requiring that  (\ref{sys:app-1st-varia}) vanishes for all $\bar \Sigma$ yields
\begin{align}
- 2 F_\gamma + V_A + V_B 
&- \frac{\Lambda_A}{R} \left\{
\left[ X\left(\sa \right) - X_C \right]  \,X'\left(\sa \right)+   
 \left[ Y\left(\sa \right) - Y_C \right]  \,Y'\left(\sa \right)   
 \right\} \nonumber \\
&+ \frac{\Lambda_B}{R} \left\{
\left[ X\left(\sb \right) - X_C \right]  \,X'\left(\sb \right)+   
 \left[ Y\left(\sb \right) - Y_C \right]  \,Y'\left(\sb \right)   
 \right\}  =0
 \label{equa:app-intensity-jumps}
\end{align}
In summary the equilibrium of the beam is governed by the system
\begin{subequations}
\label{sys:app-equil-final}
\begin{align}
X'(S)&= \cos \theta \\
Y'(S)&= \sin \theta \\
EI \theta'(S) &= M \\
M'(S) &= N_x \sin \theta - N_y \cos \theta \\
N_x'(S) &= \chi  \, \frac{\partial V}{\partial X} + \delta(S-S_A) \, \Lambda_A \, \frac{ X_A - X_C }{R}
+ \delta(S-S_B) \, \Lambda_B \, \frac{ X_B - X_C }{R}\\ 
N_y'(S) &= \chi \, \frac{\partial V}{\partial Y} + \delta(S-S_A) \, \Lambda_A \,  \frac{ Y_A - Y_C }{R}
+ \delta(S-S_B) \, \Lambda_B \, \frac{ Y_B - Y_C }{R}
\end{align}
\end{subequations}
with $\chi=1$ for $S \in \left[S_A ; S_B \right]$ and $\chi=0$ otherwise, and where $\delta(S-S_\star)$ is the Dirac distribution centered on $S=S_\star$ and $X_{A,B} = X(\frac{L}{2} \pm \Sigma)$ and $Y_{A,B} = Y(\frac{L}{2} \pm \Sigma)$.

\section{Incipient post-buckling regime} \label{section:weakly-non-lin}
%
%
%
%
%
%
%
%
%
We here focus on configurations on path $Ax$. With regard to the shifted arc-length variable $\hs=s-\ell/2$, introduced in Section  \ref{section:buckling}, the variables have the following symmetries:
\begin{subequations}
\label{symmetriesAx}
\begin{align}
x(-\hs) = 2 \, x_C - x(\hs) \, , \quad y(-\hs) = y(\hs) \, , \quad \theta(-\hs) = - \theta(\hs) \\
m(-\hs) = m(\hs) \, , \quad n_x(-\hs) = n_x(\hs) \, , \quad n_y(-\hs) = - n_y(\hs)
\end{align}
\end{subequations}
The variable $n_y(\hs)$, an odd function of $\hs$, has also to verify (\ref{eq:A24}), which reads $n_y(\hs=-\sigma)=n_y(\hs=\sigma)$. Consequently $n_y(\pm \sigma)=0$ and, as $n_y(\hs)$ is constant for $|\hs| > \sigma$, we have that $n_y(\hs) \equiv 0$, $\forall |\hs| > \sigma$.
Moreover, in the limit where the barrier potential tends to zero, $v_0 \to 0$\footnote{the repulsion from the disk is only important for configurations in the deep post-buckling regime}, $n_y'(\hs) \equiv 0$ inside the disk.
Being an odd function, $n_y(\hs)$ is then such that
\begin{equation}
n_y(\hs) \equiv 0 \quad \forall \hs
\end{equation}
The consequence is that the force jumps (\ref{sys:app-saut-forces-y}) for $n_y(\hs)$ at the entry $\hs=-\sigma$ and exit $\hs=+\sigma$ of the disk are zero: $\lambda_A \, [y_A - y_C] = 0 = \lambda_B \, [y_B-y_C]$. We discard the cases $\lambda_A=0$ and $\lambda_B=0$ for which there would not be any meniscus force at all, and conclude
\begin{equation}
y_A = y_C = y_B
\label{E0}
\end{equation}
Conditions (\ref{equa:A4b}) and (\ref{equa:A4c}) now read $(x_A-x_C)^2=1/4$ and $(x_B-x_C)^2=1/4$ which yields
\begin{equation}
x_A=x_C -1/2 \mbox{ and } x_B=x_C +1/2 \label{E1}
\end{equation}
As $n_x(-\hs) = n_x(\hs)$, the force jumps (\ref{sys:app-saut-forces-x}) imply that $\lambda_A \, [x_A-x_C]+\lambda_B \, [x_B-x_C]=0$. Using (\ref{E1}), we conclude that
\begin{equation}
\lambda_A = \lambda_B \label{E2}
\end{equation}
Finally, using the global force balance (\ref{equa:app-intensity-jumps}), (\ref{E0}), (\ref{E1}), and (\ref{E2}), and still in the limit $v_0 \to 0$, we obtain
\begin{equation}
f_\gamma = \lambda_B \,\cos \theta_B
\end{equation}
Taking advantage of the symmetries (\ref{symmetriesAx}), we now rewrite the boundary-value problem (\ref{sys:equilibre-adim}), (\ref{equa:initcnd}), (\ref{equa:finalbcnd}) for the interval $\hs \in [0 , \ell/2]$. Using the external tension $t$, we write the $x$-component of the force as $n_x(\hs) \equiv t$, $\forall \hs \in [\sigma , \ell/2]$ and using the jumps at $\hs=\sigma$ we obtain $n_x(\hs) \equiv t-\lambda < 0$, $\forall \hs \in (0 , \sigma)$ where we note $\lambda=\lambda_A=\lambda_B$.
We therefore have to solve 
\begin{subequations}
\label{B7}
\begin{align}
\d \xin / \d\hs = \cos \thin \, , \quad \d \thin / \d\hs = \Min \, , \quad \d \Min / \d\hs = - (\lambda - t) \sin \thin  
& \quad \mbox{ for } \hs \in [0,\sigma]\\
\d \xout / \d\hs = \cos \thout \, , \quad \d \thout / \d\hs = \Mout \, , \quad \d \Mout / \d\hs = - (\lambda - t) \sin \thout 
& \quad \mbox{ for } \hs \in [\sigma,\ell/2]
\end{align}
\end{subequations}
These are 6 differential equations with 2 unknowns parameters $\lambda$ and $\sigma$. Boundary conditions are
\begin{equation}
\xin(0) = x_C   \, , \quad 
\thin(0) = 0   \, , \quad 
\xin(\sigma) = x_C + 1/2  \, , \quad 
f_\gamma = \lambda \, \cos \thin(\sigma)  \, , \quad
\thout(\ell/2) =0
\label{E6}
\end{equation}
and matching conditions are
\begin{equation}
\xout(\sigma) = \xin(\sigma) \, , \quad \thout(\sigma) = \thin(\sigma) \, , \quad \Mout(\sigma) = \Min(\sigma)
\label{B9}
\end{equation}
Hence, for each given value of $t$, $f_\gamma$, $\ell$, the 8 boundary and matching conditions define a well-posed problem for equations (\ref{B7}).
For simplicity reasons, we replace the last condition in (\ref{E6}) by $\thout(+ \infty) = 0$, that is we work in the large $\ell$ limit. We look for a small amplitude solution to this boundary-value problem, that is we develop each variable in power of $\e$, where $\e$ is a small parameter.
As buckling happens through a pitchfork bifurcation, two symmetric ($\e > 0$, where the beam is buckled upward, and $\e<0$, where the beam is buckled downward) branches emerge from the $\e=0$ buckling point.
Taking advantage of this symmetry we introduce the following series
\begin{align}
\xio(\hs) & = x_C + \hs + \e^2 \, \xio_2(\hs) + O(\e^4) \label{B10} \\
\thio(\hs) & = \e \, \thio_1(\hs) + \e^3 \, \thio_3(\hs) + O(\e^5) \\
\Mio(\hs) & = \e \, \Mio_1(\hs) + \e^3 \, \Mio_3(\hs) + O(\e^5) \\
t & = t_0 + \e^2 \, t_2 + O(\e^4) \\
\lambda & = f_\gamma + \e^2 \, \lambda_2 + O(\e^4) \\
\sigma &= 1/2 + \e^2 \, \sigma_2 + O(\e^4) 
\end{align}

\begin{figure}[th]
\begin{center}
\includegraphics[width=.5\columnwidth]{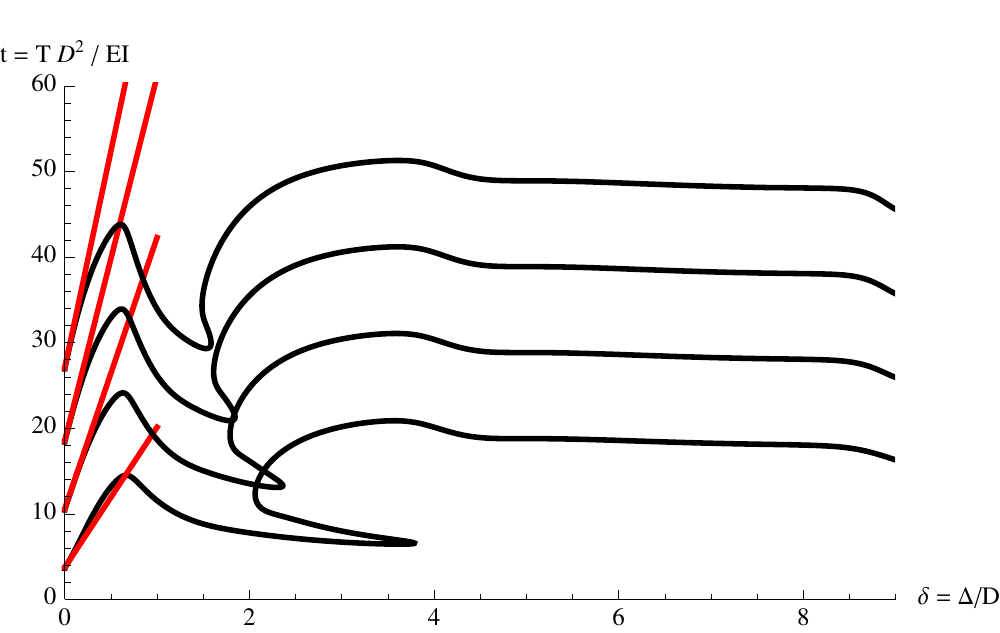}
\caption{Paths $Ax$ for $f_\gamma=20,30,40$ and $50$ with their slope at the origin given by (\ref{equa:slope-origin})}
\label{fig:fnl}
\end{center}
\end{figure}

Solving the problem (\ref{B7})-(\ref{B9}), we find at order $\e^1$ that
\begin{align}
\thin_1(\hs) & = \sin \left(\hs \R \right) \\
\thout_1(\hs) & = e^{\frac{\sqrt{t_0}}{2}} \, \sin \left( \frac{ \R}{2} \right) \, e^{- \hs \sqrt{t_0}}
\end{align}
and that $t_0$ is solution to the equation
\begin{equation}
\R \, \cos \left( \frac{\R}{2} \right) + \sqrt{t_0} \,  \sin \left( \frac{\R}{2} \right) = 0
\end{equation}
which is (\ref{equa:buckling-curve-approx-1}).
At order $\e^2$, we find 
\begin{align}
\lambda_2 & = \frac{1}{2} \, f_\gamma \, \sin^2 \left( \frac{\R}{2}\right) \\
\sigma_2 & = \frac{\R - \sin \left( \R \right) }{8 \R} \\
\xin_2(\hs) &= \frac{\sin \left( 2 \hs \R \right) - 2 \, \hs \, \R}{8 \R} \\
\xout_2(\hs) &= \frac{\sin \left( \R \right) }{8\R} -\frac{1}{8} + \frac{\sin^2 \left(\frac{\R}{2}\right) \, \left( e^{(1-2\hs)\sqrt{t_0}}-1\right)}{4\sqrt{t_0}} \label{B22}
\end{align}
At order $\e^3$, we find $\thin_3(\hs)$, $\Min_3(\hs)$, $\thout(\hs)$, and $\Mout(\hs)$ and from their matching conditions (\ref{B9}), we obtain 
\begin{equation}
t_2= \frac{f_\gamma^2 \, (4 + 5 \sqrt{t_0}) - f_\gamma  \, t_0 \, (2 + 3 \sqrt{t_0}) + 2 t_0^2 
-  2 f_\gamma \left[ f_\gamma (\sqrt{t_0}-2) + 4  t_0 \right]  \cos(\R)}
{8 f_\gamma \, (2 + \sqrt{t_0})}
\end{equation}
We now compute the end-shortening $\delta = \ell - x(s=\ell) = \ell - 2  \left[\xout(\hs=\ell/2)-x_C \right]$, still in the limit where $\ell \to +\infty$. Using (\ref{B10}) and (\ref{B22}) we find $\delta= \e^2 \delta_2$ with
\begin{align}
\delta_2= \frac{1}{4} + \frac{\sin^2 \left( \frac{\R}{2}\right)}{2\sqrt{t_0}} - \frac{\sin \R}{4\R}
\end{align}

Finally we write
\begin{equation}
t = t_0 + \e^2 t_2 = t_0 + \frac{t_2}{\delta_2} \delta
\label{equa:slope-origin}
\end{equation}
We plot in Figure~\ref{fig:fnl} paths $Ax$ for $f_\gamma=20,30,40$ and $50$ and the straight lines given by (\ref{equa:slope-origin}).

\bibliographystyle{elsarticle-num-names}
\bibliography{windlass2}
%
%
%
%
%
%
%
\end{document}